\documentclass[letterpaper,journal,final]{IEEEtran}
\IEEEoverridecommandlockouts

\newcommand{\subparagraph}{}

\usepackage{graphicx,psfrag,dblfloatfix}
\usepackage{amsmath,amssymb,amsfonts,mathrsfs}
\usepackage{color,epstopdf}
\usepackage{algorithmic}
\usepackage{enumerate} 
\usepackage{subfigure}
\usepackage{float}
\usepackage{flushend}
\usepackage{bbold}
\usepackage{dsfont}
\usepackage{mathrsfs}
\usepackage{relsize}

\newtheorem{rem}{Remark}
\newtheorem{assumption}{Assumption}
\newtheorem{definition}{Definition}
\newtheorem{theorem}{Theorem}
\newtheorem{lem}{Lemma}
\newtheorem{corollary}{Corollary}

\newcommand{\bbm}{\begin{bmatrix}}
\newcommand{\ebm}{\end{bmatrix}}

\def\qedp{\hspace*{\fill}~{\tiny $\blacksquare$}}
\def\be{\begin{equation}}
\def\ee{\end{equation}}
\def\ba{\begin{array}}
\def\ea{\end{array}}
\def\eqa{\begin{eqnarray}}
\def\eqe{\end{eqnarray}}

\definecolor{darkgreen}{rgb}{0.0, 0.55, 0.0}
\definecolor{amaranth}{rgb}{0.9, 0.17, 0.31}
\usepackage[prependcaption,colorinlistoftodos]{todonotes}

\begin{document}

\title{
Formulas for Data-driven Control: Stabilization, Optimality and Robustness
}


\author{{C. De Persis} and P. Tesi 
\thanks{
C. De Persis is with ENTEG and 
the J.C.~Willems Center for Systems and 
Control, 
University of Groningen, 9747 AG Groningen, The Netherlands.
Email: {\tt\small c.de.persis@rug.nl}.
P. Tesi is with DINFO, University of Florence, 50139 Firenze, Italy 
E-mail: {\tt\small pietro.tesi@unifi.it}.}
}

\maketitle
\begin{abstract}
In a paper by Willems and coauthors it was shown that 
persistently exciting data can be used to represent the 
input-output behavior of a linear system. Based on this 
fundamental result, we derive a parametrization of linear 
feedback systems that paves the way to solve important 
control problems using data-dependent Linear Matrix Inequalities only. 
The result is remarkable in that no explicit system's matrices identification is required. 
The examples of control problems we solve include the 
state and output feedback stabilization, and the linear quadratic regulation problem. 
We also discuss robustness to noise-corrupted measurements 
and show how the approach can be used to stabilize unstable equilibria of 
nonlinear systems.
\end{abstract}

\section{Introduction}

\IEEEPARstart{L}{earning} from data is essential to every area of science.
It is the core of statistics and artificial intelligence, and is becoming 
ever more prevalent also in the engineering domain. 
Control engineering is one of the domains where learning 
from data is now considered as a prime issue.

Learning from data is actually not novel in control theory.
System identification \cite{Ljung1987} is one of the major developments of 
this paradigm, where modeling based on first principles is replaced 
by data-driven learning algorithms. Prediction error, maximum likelihood
as well as subspace methods \cite{verhaegen2007filtering} are all data-driven techniques 
which can be now regarded as standard for what concerns modeling.
The learning-from-data paradigm has been widely pursued also
for control design purposes. 
A main question is how to design control systems 
directly from process data with no intermediate system identification step. 
Besides their theoretical value, answers to this question could 
have a major practical impact especially in those situations
where identifying a process model can be difficult and time consuming,
for instance when data are affected by noise 
or in the presence of nonlinear dynamics.
Despite many developments in this area, 
data-driven control is not yet well understood even if we restrict 
the attention to linear dynamics, which contrasts the achievements obtained in system
identification. A major challenge is how to incorporate data-dependent stability 
and performance requirements in the control design procedure. \smallskip

\emph{Literature review}

Contributions to data-driven control can be traced back to the
pioneering work by Ziegler and Nichols \cite{Ziegler1942},
{direct} adaptive control
\cite{Astrom1989} and neural networks \cite{werbos1989} theories.
Since then, many techniques have been developed 
under the heading \emph{data-driven} and \emph{model-free} control. We mention unfalsified 
control theory \cite{safo1995}, iterative feedback tuning \cite{Hjalmarsson1998},
and virtual reference feedback tuning \cite{Campi2002}.
This topic is now attracting more and more researchers,
with problems ranging
from PID-like control \cite{Fliess2013} to 
model reference control and output tracking
\cite{Karimi2,Karimi4,Campestrini2017,D2IBC,DFK}, 
predictive \cite{Salvador2018,Coulson2018}, robust \cite{Dai2018}
and optimal control \cite{Markovsky2007,Vamvoudakis2010,Jiang2012,
Pang2018,Mukherjee2018,Goncalves2019,Baggio2019}, 
the latter being one of the most frequently considered problems.
The corresponding techniques are also quite varied, ranging 
from dynamics programming to optimization techniques and algebraic methods.
These contributions also differ with respect to how learning is approached.
Some methods only use a batch of process data meaning that learning is performed
off-line, while other methods are iterative and require multiple on-line experiments.
We refer the reader to \cite{Bazanella2011,Hou2013} for more 
references on data-driven control methods.

\smallskip

\emph{Willems \emph{et al.}'s fundamental lemma and paper contribution}

A central question in data-driven control is how
to replace process models with data. For linear systems,
there is actually a fundamental result which answers this question,
proposed by Willems \emph{et al.} \cite{willems2005note}.
Roughly, this result stipulates that the whole set of trajectories that a linear 
system can generate can be represented by a finite set of system trajectories
provided that such trajectories come from sufficiently excited dynamics. 
While this result has been (more or less explicitly) used for data-driven control design
\cite{Coulson2018,Markovsky2007,Markovsky2008,Park2009,Maupong2017},
certain implications of the so-called \emph{Willems \emph{et al.}'s fundamental lemma}
seems not fully exploited.

In this paper, we first revisit Willems \emph{et al.}'s fundamental lemma,
originally cast in the behavioral framework, through classic 
state-space descriptions (Lemma \ref{willems.fundamental}).  
Next, we show that this result can be used to get a 
data-dependent representation of the open-loop and closed-loop
dynamics under a feedback interconnection. The first result
(Theorem \ref{prop:sys.identif}) indicates that
the parametrization that emerges from the fundamental lemma 
is in fact the solution to a classic least-squares problem, and has 
clear connections with the so-called Dynamic Mode Decomposition
\cite{Proctor2014}. The second result (Theorem \ref{prop:data-driven-stab}) 
is even more interesting as it provides a data-based representation 
of the closed-loop system transition matrix, where the 
controller is itself parametrized through data.

Theorem \ref{prop:data-driven-stab} turns out to have surprisingly straightforward, 
yet profound, implications for control design. We discuss this fact in 
Section \ref{sec:direct-data-driven}. 
The main point is that the parametrization
provided in Theorem \ref{prop:data-driven-stab} can be naturally 
related to the classic Lyapunov stability inequalities. This makes it possible 
to cast the problem of designing state-feedback controllers in terms of 
a simple Linear Matrix Inequality (LMI) \cite{scherer2004} (Theorem \ref{prop:data-driven-design}). 
In Theorem \ref{prop:data-driven-lqr},
the same arguments are used to solve a linear quadratic regulation
problem through convex optimization. 
A remarkable feature of these results is that: 
(i) no parametric model of system is identified; 
(ii) stability guarantees 
come with a finite (computable) number of data points.
Theorems \ref{prop:data-driven-design} and \ref{prop:data-driven-lqr}
should be understood as \emph{examples} of how the parametrization given in
Theorem \ref{prop:data-driven-stab} can be used to approach the direct design of control laws from data.
In fact, LMIs have proven their effectiveness in a variety of control design
problems \cite{scherer2004}, and we are confident that the same arguments
can be used for approaching other, more complex, design problems such as
$H_\infty$ control and quadratic stabilization \cite{scherer2004}. 
In Section \ref{sec:direct-data-driven_rob}, 
we further exemplify the merits of the proposed approach
by considering the problem of designing stabilizing controllers 
when data are corrupted by noise
(Theorem \ref{prop:noise_stability}), 
as well as the problem of stabilizing an unstable equilibrium of 
a nonlinear system (Theorem \ref{prop:nonlinear_stability}), 
both situations where identification can be challenging.
 The main derivations are given for state feedback. 
 The case of output feedback (Theorem \ref{output-feedback-controller})
is discussed in Section \ref{sec:of-siso}. Concluding remarks are given 
in Section \ref{sec:conclusion}.

\subsection{Notation}
Given a signal $z:
\mathbb{Z}\to \mathbb{R}^\sigma$, we denote by $z_{[k, k+T]}$, where
$k\in \mathbb{Z}$, $T\in \mathbb{N}$, the restriction in vectorized form 
of $z$ to the interval $[k, k+T]\cap \mathbb{Z}$, namely
\[
z_{[k, k+T]}=
\bbm
z(k)\\
\vdots 
\\
z( k+T)
\ebm.
\]
When the signal is not restricted to an interval 
then it is simply denoted by its symbol, say $z$. 
To avoid notational burden, we use
$z_{[k, k+T]}$ also to denote the sequence $\{z(k),\ldots,z(k+T)\}$.
For the same reason, we simply write $[k, k+T]$ to denote the discrete 
interval $[k, k+T]\cap \mathbb{Z}$.

We denote the Hankel matrix associated to $z$ as
\[
Z_{i,t,N}
=
\begin{bmatrix}
z(i) & z(i+1) & \cdots & z(i+N-1)\\
z(i+1) & z(i+2) & \cdots & z(i+N)\\
\vdots & \vdots & \ddots & \vdots\\
z(i+t-1) & z(i+t)  & \cdots & z(i+t+N-2)\\
\end{bmatrix}
\]
where $i \in \mathbb{Z}$ and $t, N\in \mathbb{N}$.
The first subscript denotes the time at which the first sample of the signal is taken, 
the second one the number of samples per each column, 
and the last one the number of signal samples per each row. 
Sometimes, if $t=1$, noting that the matrix
$Z_{i,t,N}$ has only one block row, we simply write
\[
Z_{i,N}=
\begin{bmatrix}
z(i) & z(i+1) & \cdots & z(i+N-1)\\
\end{bmatrix}.
\]

\section{
Persistence of excitation 
and
Willems \emph{et al.}'s fundamental lemma
}
\label{sec:fund.lemma}

In this section, we revisit the main result in \cite{willems2005note} and state a few auxiliary results inspired by subspace identification \cite{verhaegen2007filtering}, which will be useful throughout the paper.  

For the sake of simplicity, throughout the paper  we consider a 
controllable and observable
discrete-time linear system 
\begin{subequations}\label{lin.sys}
\begin{align}
x(k+1) =  A x(k) + B u(k)\label{lin.sys.1}\\
y(k) =  C x(k) + D u(k)\label{lin.sys.2}
\end{align}
\end{subequations}
where $x\in \mathbb{R}^n, u\in \mathbb{R}^m$ and $y\in \mathbb{R}^p$.
The system input-output response of a
over $[0, t-1]$ can be expressed as
\be
\label{fund.eqn}
\bbm
u_{[0, t-1]}\\
y_{[0, t-1]}
\ebm
=
\begin{bmatrix}
I_t & \mathbb{0}_{tm\times n}\\
\hline
\mathcal{T}_t & \mathcal{O}_t
\end{bmatrix}
\bbm
u_{[0, t-1]}\\
\hline
x_0
\ebm
\ee
where $x_0$ is the system initial state, and where
\begin{eqnarray*} \label{toep.obs.matrx}
&& \mathcal{T}_t := 
\begin{bmatrix}
D & 0 & 0 & \cdots & 0 \\
CB & D & 0 &\cdots & 0\\
CAB & CB & D & \cdots & 0\\
\vdots &\vdots &\vdots &\ddots & \vdots \\
CA^{t-2}B & C A^{t-3} B & C A^{t-4} B & \cdots & D\\
\end{bmatrix}, \nonumber
\\
&& \mathcal{O}_t := 
\begin{bmatrix}
C\\
CA\\
\vdots\\
C A^{t-1}
\end{bmatrix}
\end{eqnarray*}
are the Toeplitz and observability matrices of order $t$. 

Let now $u_{d, [0, T-1]}$ and $y_{d, [0, T-1]}$ 
be the
input-output data of the system 
collected during an experiment,
and let
\begin{equation} \label{eq.Ht.0}
\begin{bmatrix}
U_{0,t,T-t+1}\\
\hline
Y_{0,t,T-t+1}
\end{bmatrix} :=
\begin{bmatrix}
u_d(0) & u_d(1) & \cdots & u_d(T-t)\\
u_d(1) & u_d(2) & \cdots & u_d(T-t+1)\\
\vdots & \vdots & \ddots & \vdots\\
u_d(t-1) & u_d(t) & \cdots & u_d(T-1)\\
\hline
y_d(0) & y_d(1) & \cdots & y_d(T-t)\\
y_d(1) & y_d(2) & \cdots & y_d(T-t+1)\\
\vdots & \vdots & \ddots & \vdots\\
y_d(t-1) & y_d(t) & \cdots & y_d(T-1)\\ 
\end{bmatrix} 
\smallskip
\end{equation}
be the corresponding Hankel matrix.
Similarly to \eqref{fund.eqn}, we can write
\be\label{eq.Ht}
\begin{bmatrix}
U_{0,t,T-t+1}\\
\hline
Y_{0,t,T-t+1}
\end{bmatrix}=
\begin{bmatrix}
I_{tm} & \mathbb{0}_{tm\times n}\\
\hline
\mathcal{T}_t & \mathcal{O}_t
\end{bmatrix}
\begin{bmatrix}
U_{0, t,T-t+1} \\
\hline
X_{0, T-t+1}
\end{bmatrix}
\ee
where 
\[
X_{0, T-t+1}=
\begin{bmatrix}
x_{d}(0) & x_{d}(1)  & \dots & x_{d}(T-t)
\end{bmatrix}
\]
and $x_{d}(i)$
are the state samples. For $u_d$, $y_d$, and $x_d$, we use
the subscript $d$ so as to emphasize that these are the sample
data collected from the system during some experiment.

\subsection{
Persistently exciting data and the fundamental lemma}
\label{subsec:fund.lemma}

Throughout the paper, having the rank condition 
\be\label{rank.condition}
{\rm rank}
\begin{bmatrix}
U_{0, t,T-t+1} \\
\hline
X_{0, T-t+1}
\end{bmatrix}
=
n+ tm
\ee
satisfied plays an important role. 
As we will see, a condition of this type in fact ensures that the data
encode all the information for the direct design of control laws.
A fundamental property established in \cite{willems2005note}
is that it is possible to guarantee \eqref{rank.condition} 
when the input is sufficient exciting. 
We first recall the notion of persistency of excitation.

\smallskip
\begin{definition}
{\rm \cite{willems2005note}}
The signal $z_{[0, T-1]}\,\in \mathbb R^\sigma$ 
is persistently exciting of order $L$ if the matrix 
\[
Z_{0,L,T-L+1}
=
\begin{bmatrix}
z(0) & z(1) & \cdots & z(T-L)\\
z(1) & z(2) & \cdots & z(T-L+1)\\
\vdots &\vdots &\ddots & \vdots \\
z(L-1) & z(L) & \cdots & z(	T-1)\\
\end{bmatrix}
\]
has full rank $\sigma L$. \qedp
\end{definition}
\smallskip

For a signal $z$
to be persistently exciting of order $L$, it must be sufficiently long, namely 
$T\ge (\sigma+1)L-1$.  
We now state two results which are key
for the developments of the paper.

\smallskip
\begin{lem}\label{lem:willems}
{\rm \cite[Corollary 2]{willems2005note}}
Consider  system \eqref{lin.sys.1}.
If the input $u_{d, [0, T-1]}$ 
is persistently exciting of order $n+t$, then condition \eqref{rank.condition} holds. \qedp
\end{lem} 
\smallskip

\begin{lem}\label{willems.fundamental}
{\rm \cite[Theorem 1]{willems2005note}}
Consider  system \eqref{lin.sys}.
Then the following holds:
\begin{enumerate}[(i)]
\item 
If $u_{d, [0, T-1]}$ is persistently exciting of order $n+t$, then any $t$-long 
input/output trajectory of system \eqref{lin.sys} can be expressed as 
\[
\bbm
u_{[0, t-1]}\\
y_{[0, t-1]}
\ebm
=
\begin{bmatrix}
U_{0,t,T-t+1}\\
\hline
Y_{0,t,T-t+1}
\end{bmatrix} g
\]
where $g\in \mathbb{R}^{T-t+1}$.
\item 
Any linear combination of the columns of the matrix in \eqref{eq.Ht.0}, that is
\[
\begin{bmatrix}
U_{0,t,T-t+1}\\
\hline
Y_{0,t,T-t+1}
\end{bmatrix}g,
\]
is a $t$-long input/output trajectory of \eqref{lin.sys}.
\end{enumerate}
\end{lem} \smallskip

{\it Proof.} See the Appendix. \qedp \smallskip

Lemma \ref{lem:willems} shows that if $T$ is taken sufficiently large
then \eqref{rank.condition} turns out to be satisfied, and 
this makes it possible to represent
any input/output trajectory of the system as a linear combination 
of collected input/output data. 
This is  the key property that enables one to replace a parametric description
of the system with data.
Lemma \ref{willems.fundamental} has been originally proven in 
\cite[Theorem 1]{willems2005note}
using the behavioral language, and it was later referred to in \cite{Markovsky2005} as the
\emph{fundamental lemma} to describe a linear system 
through a finite collection of its input/output data. Here, for making the paper as self-contained as possible, we gave a proof of this result 
using state-space descriptions, as they will recur often in the reminder of this paper. 

\section{Data-based system representations} \label{sec:db_reprsentation}

Lemma \ref{willems.fundamental} allows us to 
get a data-dependent representation of the open-loop and closed-loop
dynamics of system \eqref{lin.sys.1}. The first result (Theorem \ref{prop:sys.identif})
is a covert system identification result where, however, the role of
Lemma \ref{willems.fundamental} is emphasized, and which draws 
connections with the so-called Dynamic Mode Decomposition
\cite{Proctor2014}. Theorem \ref{prop:data-driven-stab} shows instead
how one can parametrize feedback interconnections just by using data. This result will
be key later on for deriving control design methods 
that avoid the need to identify a parametric model of the system to be controlled.

Consider a persistently exciting input sequence $u_{d, [0, T-1]}$ 
of order $t+n$ with $t=1$.
Notice that the only requirement on $T$ is that 
$T\ge (m+1) n +m$,  
which is necessary for the persistence of excitation condition to hold. 
By Lemma \ref{lem:willems}, 
\be\label{Ud}
{\rm rank} \begin{bmatrix}
U_{0,1,T} \\
\hline
X_{0, T}
\end{bmatrix} = n+m.
\ee
From now on, we will directly refer to condition \eqref{Ud},
bearing in mind that this condition requires 
persistently exciting inputs of order 
$n+1$. 
Before proceeding, we point out that condition \eqref{Ud} can always be directly
assessed when the state of the system is accessible. 
When instead only 
input/output data are accessible, condition \eqref{Ud} cannot be directly assessed.
Nonetheless, thanks to Lemma \ref{lem:willems} this condition can always be enforced 
by applying an exciting input signal of a sufficiently high order
-- for a discussion on the types of persistently exciting 
signals the reader is referred to \cite[Section 10]{verhaegen2007filtering}.
We will further elaborate on this point in Section \ref{sec:of-siso} where we also
give an alternative explicitly verifiable condition for the case where only input/output data 
of the system are accessible.

\subsection{Data-based open-loop representation}

The next result gives a data-based representation of a linear system and
emphasizes the key role of Lemma \ref{willems.fundamental}. \smallskip

\begin{theorem}\label{prop:sys.identif}
Let condition \eqref{Ud} hold. Then system \eqref{lin.sys.1} 
has the following equivalent representation
\be\label{data-model}
x(k+1)
= 
X_{1,T} 
\begin{bmatrix}
U_{0,1,T} \\
\hline
X_{0, T}
\end{bmatrix}
^\dag
\begin{bmatrix}
u(k)\\
x(k)
\end{bmatrix}
\ee
where 
\[
X_{1,T}=\begin{bmatrix} x_{d}(1) & {x_{d}}(2)  & \ldots & {x_{d}}(T)
\end{bmatrix}
\] 
and $\dag$ denotes the right inverse.
\end{theorem} \smallskip

\emph{Proof.}  See the Appendix. \qedp \smallskip

Theorem \ref{prop:sys.identif} is an identification type of 
result where the role of Lemma \ref{willems.fundamental} is made explicit. In fact,  
noting that
\begin{eqnarray}
X_{1, T} =
\begin{bmatrix}
B & A
\end{bmatrix}
\begin{bmatrix}
U_{0,1,T} \\
\hline
X_{0, T}
\end{bmatrix}
\end{eqnarray}
it follows immediately that 
\begin{eqnarray} \label{sol.least.square}
\begin{bmatrix}
B & A
\end{bmatrix}
=
X_{1, T}
\begin{bmatrix}
U_{0,1,T} \\
\hline
X_{0, T}
\end{bmatrix}^\dag .
\end{eqnarray}
In particular, the right-hand side of the above identity
is simply the minimizer of the least-square problem 
\cite[Exercise 9.5]{verhaegen2007filtering} 
\begin{eqnarray} \label{least.square}
{\min}_{[B \,\, A]} 
\left\| X_{1, T} - 
\begin{bmatrix} B & A\end{bmatrix} 
\begin{bmatrix}
U_{0,1,T} \\
\hline
X_{0, T}
\end{bmatrix}
\right\|_{{\rm F}}
\end{eqnarray}
where $\|\cdot\|_{F}$ is the Frobenius norm. The representation given in 
Theorem \ref{prop:sys.identif} can be thus interpreted as the solution of 
a least-square problem.

It is also interesting to observe that Theorem \ref{prop:sys.identif}
shows clear connections between Willems \emph{et al.}'s fundamental lemma 
and the Dynamic Mode Decomposition \cite{Proctor2014},
a numerical procedure for recovering state and control matrices
of a linear system from its trajectories. In fact, 
by performing a singular value decomposition 
\[
\begin{bmatrix}
U_{0,1,T} \\
\hline
X_{0, T}
\end{bmatrix} = 
U_1 \Sigma V_1^\top ,
\]
it readily follows that \eqref{sol.least.square} can be rewritten 
as $X_{1, T} V_1 \Sigma^{-1} U_1^\top$ \cite[Section 2.6]{verhaegen2007filtering},
which is the basic solution described in \cite[Section III-B]{Proctor2014}
for recovering the matrices $A$ and $B$ of a linear system from its
trajectories.

\subsection{Data-based closed-loop representation}
\label{sec:data-driven-parametrization} 
 
We now exploit Lemma \ref{willems.fundamental} 
to derive a parametrization of system \eqref{lin.sys.1}  in closed-loop with a 
state-feedback law $u=K x$. 
We give here a proof of this result 
since the arguments we use will often recur in 
the next sections.

\smallskip
\begin{theorem}\label{prop:data-driven-stab}
Let condition \eqref{Ud} hold. Then system \eqref{lin.sys.1} 
in closed-loop with a state feedback $u=Kx$ has the following equivalent representation 
\be\label{closed.loop.param}
x(k+1) =X_{1, T} G_K x(k)
\ee
where $G_K$ is a $T\times n$ matrix satisfying 
\be\label{compute.G_K}
\begin{bmatrix}
K \\ I_n
\end{bmatrix}
= 
\begin{bmatrix}
U_{0,1,T} \\
\hline
X_{0, T}
\end{bmatrix}
G_K .
\ee
In particular
\be\label{data-control}
u(k)= U_{0,1,T} G_K x(k).
\ee
\end{theorem}

{\it Proof.} 
By the Rouch\'{e}-Capelli theorem, there exists a 
$T\times n$ matrix $G_K$ such that \eqref{compute.G_K} holds. 
Hence, 
{\setlength\arraycolsep{2.3pt}
\begin{eqnarray} 
\label{important3}
\ba{rl}
A+BK 
=&
\begin{bmatrix}
B & A
\end{bmatrix}
\begin{bmatrix}
K \\ I_n
\end{bmatrix}
= \begin{bmatrix}
B & A
\end{bmatrix}
\begin{bmatrix}
U_{0,1,T} \\
\hline
X_{0, T}
\end{bmatrix}
G_K 
\\
= & 
X_{1, T}
G_K 
\ea
\end{eqnarray}}%
In particular, the first identity in \eqref{compute.G_K} gives \eqref{data-control}.
\qedp

\subsection{From indirect to direct data-driven control}

Obviously, Theorem \ref{prop:sys.identif} already provides a way 
for designing controllers from data, at least when the state of the system
to be controlled is fully accessible. However, this approach 
is basically equivalent to a model-based approach where 
the system matrices $A$ and $B$ are first reconstructed using a {collection} 
of sample trajectories. A {crucial} observation that emerges from
Theorem \ref{prop:data-driven-stab} is that also the controller $K$
can be parametrized through data via \eqref{compute.G_K}.
Thus for design purposes one can regard $G_K$ as a \emph{decision
variable}, and search for the matrix $G_K$ that guarantees   
stability and performance specifications. 
In fact, as long as $G_K$ satisfies the condition {$X_{0,T} G_K = I_n$} in
\eqref{compute.G_K} we are ensured that $X_{1,T} G_K$ provides an equivalent 
representation of the closed-loop matrix $A+BK$ with feedback matrix
$K = U_{0,1,T} G_K$. As shown in the next section, this enable 
design procedures that avoid the need to identify a parametric model of the system.

We point out that Theorem \ref{prop:data-driven-stab}
already gives an identification-free method for \emph{checking} whether 
a candidate controller $K$ is stabilizing or not. In fact, given $K$, any solution 
$G_K$ to \eqref{compute.G_K} is such that $X_{1,T} G_K = A+BK$.
One can therefore compute the eigenvalues of $X_{1,T} G_K$ to check 
whether $K$ is stabilizing or not. This method does not require
to place $K$ into feedback, in the spirit of unfalsified control theory \cite{safo1995}. 

\section{
Data-driven control design: stabilization
and optimal control}
\label{sec:direct-data-driven}

In this section, we discuss how Theorem \ref{prop:data-driven-stab} 
can be used to get identification-free design algorithms. Although the 
problems considered hereafter are all of practical relevance, we
would like to regard them as application \emph{examples} of 
Theorem \ref{prop:data-driven-stab}. In fact, we are confident that 
Theorem \ref{prop:data-driven-stab}
can be used to approach other, more complex, design problems such as
$H_\infty$ control and quadratic stabilization \cite{scherer2004}.

\subsection{
State feedback design and data-based parametrization of all stabilizing controllers}
\label{subsec:data-driven-parametrization}

By Theorem \ref{prop:data-driven-stab}, the closed-loop system
under state-feedback $u=Kx$ is such that
\begin{eqnarray}
A+BK = X_{1, T} G_K \nonumber
\end{eqnarray}
where $G_K$ satisfies \eqref{compute.G_K}.
One can therefore search for a matrix $G_K$
such that $X_{1, T} G_K$ satisfies the classic
Lyapunov stability condition. As the next result shows, it turns 
out that this problem can be actually cast in terms of a simple
Linear Matrix Inequality (LMI). \smallskip
 
\begin{theorem}\label{prop:data-driven-design}
Let condition \eqref{Ud} hold.
Then, any matrix $Q$ satisfying 
\begin{eqnarray}
\label{lmi.state.feedback}
\begin{bmatrix}
X_{0,T} \,Q &  X_{1,T} Q\\
Q^\top X_{1,T}^\top & X_{0,T} \,Q
\end{bmatrix}
\succ 0
\end{eqnarray}
is such that
\begin{eqnarray}
\label{data-control-explicit}
K = U_{0,1,T}Q (X_{0,T} Q )^{-1}
\end{eqnarray}
stabilizes system \eqref{lin.sys.1}.
{Conversely, if $K$ is a stabilizing state-feedback gain
for system \eqref{lin.sys.1}
then 
it can be written as in
\eqref{data-control-explicit}, with $Q$ solution of \eqref{lmi.state.feedback}.}
\smallskip
\end{theorem}

{\em Proof.} 
By Theorem \ref{prop:data-driven-stab}, 
\eqref{closed.loop.param} is an equivalent representation of the closed-loop system. 
Hence, for any given $K$
the closed-loop system with $u=Kx$ is asymptotically stable
if and only if there exists $P\succ 0$ such that 
\begin{eqnarray} \label{lyap.ineq}
X_{1, T} G_K P G_K^\top X_{1, T}^\top  - P\prec 0
\end{eqnarray}
where $G_K$ satisfies \eqref{compute.G_K}. 

Let $Q:=G_K P$.
Stability is thus equivalent to the existence of two matrices $Q$ and $P\succ 0$ such that
{\setlength\arraycolsep{2pt} 
\begin{eqnarray}\label{alternative}
\def\arraystretch{1.2}
\left\{
\ba{l}
X_{1, T} Q P^{-1} Q^\top X_{1, T}^\top  - P\prec  0 \\
X_{0, T} Q = P \\
U_{0,1,T}Q = KP
\ea \right.
\end{eqnarray}}%
where the two equality constraints are obtained from \eqref{compute.G_K}.
By exploiting the constraint $X_{0, T} Q = P$, stability is equivalent 
to the existence of a matrix $Q$ such that 
{\setlength\arraycolsep{2pt} 
\begin{eqnarray} \label{alternative_2}
\def\arraystretch{1.2}
\left\{
\ba{l}
X_{1, T} Q (X_{0, T} Q)^{-1} Q^\top X_{1, T}^\top  - X_{0, T} Q \prec  0 \\
X_{0, T} Q \succ 0 \\
U_{0,1,T}Q = K X_{0, T} Q
\ea \right.
\end{eqnarray}}%
From the viewpoint of design, one can thus focus on the 
two inequality constraints which correspond to \eqref{lmi.state.feedback},
while the equality constraint is satisfied a posteriori with the 
choice $K = U_{0,1,T}Q (X_{0,T} Q )^{-1}$. \qedp \smallskip

Note that in the formulation \eqref{lmi.state.feedback}
the parametrization of the closed-loop matrix $A+BK$ is given by 
$X_{1, T} Q (X_{0, T} Q)^{-1}$, that is with $G_K = Q (X_{0, T} Q)^{-1}$
which satisfies $X_{0, T} G_K = I$ corresponding to the second 
identity in \eqref{compute.G_K}. On the other hand, the constraint corresponding 
to the first identity in \eqref{compute.G_K} is guaranteed 
by the choice $K = U_{0,1,T}Q (X_{0,T} Q )^{-1}$. This is the reason why
\eqref{lmi.state.feedback} is representative of closed-loop stability 
even if no constraint like \eqref{compute.G_K} appears in the formulation \eqref{lmi.state.feedback}.
We point out that Theorem \ref{prop:data-driven-design}
characterizes the whole set of stabilizing state-feedback gains
in the sense that any stabilizing feedback gain $K$ can be 
expressed as in \eqref{data-control-explicit} for some matrix $Q$ 
satisfying \eqref{lmi.state.feedback}. 
\smallskip

\emph{Illustrative example}. As an illustrative example, consider the
discretized version of a batch reactor system \cite{Walsh2001} using a sampling time of $0.1s$,

{
\small
\begin{align} 
& \left[
\ba{c|c}
A & B
\ea
\right] = \nonumber \\
& \quad
\left[
\ba{rrrr|rr}
    1.178  &  0.001 &   0.511 &  -0.403 & 0.004 &   -0.087 \\
   -0.051 &    0.661 &   -0.011 &   0.061 &     0.467 &   0.001 \\
    0.076 &    0.335 &    0.560 &    0.382 &  0.213 &   -0.235 \\
   0 &    0.335 &   0.089 &   0.849 & 0.213 &   -0.016
\ea
\right]\,. \nonumber
\end{align}
}%
\normalsize
The system to be controlled is open-loop unstable.
The control design procedure is implemented in 
MATLAB. We generate the data with random initial conditions and by applying to each input channel 
a random input sequence of length $T=15$ by using the MATLAB command \texttt{rand}.
To solve \eqref{lmi.state.feedback} we used CVX \cite{cvx}, obtaining 
\begin{eqnarray} 
K = \left[
\ba{rrrr}
0.7610  & -1.1363 &   1.6945 &  -1.8123 \\
3.5351  &  0.4827 &   3.3014  & -2.6215
\ea
\right]\,, \nonumber
\end{eqnarray}
which stabilizes the closed-loop dynamics in agreement with 
Theorem \ref{prop:data-driven-stab}. \qedp \smallskip

\begin{rem}
\emph{(Numerical implementation)} There are other ways
to implement \eqref{lmi.state.feedback}. One of these alternatives is obtained from \eqref{alternative},  considering the first inequality, the third equality and condition $P \succ 0$, and rewriting them as 
\[
\begin{bmatrix}
P & X_{1,T} Q  \\
Q^\top X_{1,T}^\top & P
\end{bmatrix}
\succ 0, \quad X_{0,T}Q=P. 
\]
In this case the resulting stabilizing state feedback-gain 
takes the expression $K= U_{0,1,T} Q P^{-1}$. 
In the previous numerical example but also in those 
that follow we observed that a formulation like the one above is  
more stable numerically. The reason is that CVX cannot 
directly interpret \eqref{lmi.state.feedback} as a symmetric matrix
(the upper-left block is given by $X_{0,T} Q$ with non-symmetric decision variable $Q$),
and returns a warning regarding the expected outcome.
\qedp
\end{rem}
\smallskip

\begin{rem} \emph{(Design for continuous-time systems)}
Similar arguments can be used to deal with continuous-time systems. 
Given a sampling time $\Delta>0$, let
\[\ba{rl}
U_{0,1,T}= &
\bbm
u_d(0) & u_d(\Delta) & \ldots &  u_d((T-1)\Delta)
\ebm\\[2mm]
X_{0,T}=&
\bbm
x_d(0) & x_d(\Delta) & \ldots &  x_d((T-1)\Delta)
\ebm.
\ea\]
be input and state sampled trajectories.
Under condition \eqref{Ud} (note that, 
if 
the sequence $u_d(0), u_d(\Delta), \ldots$ is persistently exciting 
of order $n+1$,  then the application of the zero-order hold signal obtained from the  input samples above 
ensures condition \eqref{Ud} for the sampled-data system for generic choices of $\Delta$)
we have
$A+BK = X_{1, T} G_K$
where
\[
X_{1,T} :=
\bbm
\dot x_d(0) & \dot x_d(\Delta) & \ldots &  \dot x_d((T-1)\Delta)
\ebm.
\]
Hence, for any given $K$,
the closed-loop system with $u=Kx$ is asymptotically stable
if and only if there exists $P\succ 0$ such that 
\[\ba{rl}
X_{1, T} G_K P + P G_K^\top X_{1, T}^\top 
\prec 0,
\ea
\]
where $G_K$ satisfies \eqref{compute.G_K}. 
In full analogy with the discrete-time case, it follows that
any matrix $Q$ satisfying 
{\setlength\arraycolsep{2pt} 
\begin{eqnarray}
\label{lmi.state.feedback.ct}
\def\arraystretch{1.2}
\left\{
\ba{l}
X_{1, T}
Q + Q^\top X_{1, T}^\top 
\prec  0\\
X_{0,T} Q  \succ  0 
\ea \right.
\end{eqnarray}}%
is such that $K = U_{0,1,T}Q (X_{0,T} Q )^{-1}$
is a stabilizing feedback gain.
The main difference with respect to the case of discrete-time systems is the presence of the matrix $X_{1,T}$ that contains the  derivatives of the state at the sampling times, which are usually not available as measurements. The use of these methods in the context of continuous-time systems might require the use of filters for the approximation of derivatives \cite{garnier2003continuous,larsson2008estimation,padoan2015towards}. 
{This is left for future research.} 
We stress that even though the matrix \eqref{Ud} is built starting from input and state 
samples, the feedback gain $K = U_{0,1,T}Q (X_{0,T} Q )^{-1}$, 
where $Q$ is the solution of \eqref{lmi.state.feedback.ct}, 
stabilizes the {\em continuous-time} system, not its sampled-data model.
\qedp
\end{rem}
\smallskip

\subsection{Linear quadratic regulation}

Matrix (in)equalities similar to the one in \eqref{lmi.state.feedback} 
are recurrent in control design, with the major difference that in 
\eqref{lmi.state.feedback} only information collected from data appears, 
rather than the system matrices. Yet, these matrix inequalities can inspire 
the data-driven solution of other control problems. Important examples
are optimal control problems.

Consider the system 
{\setlength\arraycolsep{2pt} 
\begin{eqnarray} \label{eq:system_h2}
\def\arraystretch{1.3}
\begin{array}{r}
x(k+1) = Ax(k)+Bu(k)+\xi(k) \\[0.1cm]
z(k) = \left[
\begin{array}{cc}
Q_x^{1/2} & 0 \\ 0 & R^{1/2}
\end{array}
\right] \left[
\begin{array}{c}
x(k) \\ u(k)
\end{array}
\right] 
\end{array}
\end{eqnarray}}%
where $\xi$ is an external input to the system, 
and where $z$ is a performance signal of interest;
$Q_x \succeq 0$, $R \succ 0$ are weighting matrices with $(Q_x,A)$ observable.
The objective is to design a state-feedback law $u=Kx$ which renders
$A+BK$ stable and minimizes 
the $H_2$ norm of the transfer function $h :\xi \rightarrow z$ \cite[Section~4]{Chen1995},
\begin{eqnarray}
\| h \|_2 := \left[ \frac{1}{2 \pi} \int_{0}^{2 \pi} {\rm trace} 
\left( h \left(e^{j \theta}\right)^\top h \left(e^{j \theta}\right) \right) d\theta \right]^{\frac{1}{2}}.
\end{eqnarray}
This corresponds in the time domain to the $2$-norm 
of the output $z$ when impulses are applied to the input channels, and it can also
be interpreted as the mean-square deviation of $z$ when 
$\xi$ is a white process with unit covariance.
It is kwown \cite[Section~6.4]{Chen1995} that the solution to
this problem is given by the controller
\[
K = - (R+ B^\top X B)^{-1} B^\top X A
\] 
where $X$ is the unique  positive definite solution to the discrete-time algebraic Riccati (DARE) equation
{\setlength\arraycolsep{2pt}
\begin{eqnarray} \label{eq:H2_Riccati}
&& A^\top X A - X \\ 
&& \quad - (A^\top X B)(R+B^\top X B)^{-1} (B^\top X A )+Q_x = 0.
\nonumber 
\end{eqnarray} 

This problem of finding $K$ 
can be equivalently formulated as a convex program \cite{feron1992numerical,Balakrishnan2003}.
To see this, notice that 
the closed-loop system is given by
\begin{eqnarray} \label{eq:closed}
\left[
\begin{array}{c}
x(k+1) \\ z(k)
\end{array}
\right]  = 
\left[
\begin{array}{c|c}
A+BK & I \\ \hline 
\left[ \begin{array}{c} Q_x^{1/2} \\ R^{1/2} K \end{array} \right]  & 0
\end{array}
\right] \left[
\begin{array}{c}
x(k) \\ \xi(k)
\end{array}
\right] 
\end{eqnarray}
with corresponding $H_2$ norm 
\begin{eqnarray} \label{eq:h2}
\| h \|_2 = \left[ {\rm trace}  
\left(  Q_x W_c +  K^\top R K  W_c  \right) \right]^{\frac{1}{2}}
\end{eqnarray}
where $W_c$ denotes the controllability Gramian of the closed-loop system \eqref{eq:closed}, 
which satisfies
\begin{eqnarray}
(A+BK) W_c (A+BK)^\top - W_c + I = 0\nonumber
\end{eqnarray}
where $W_c \succeq I $.
The second term appearing in the {trace} function is equivalent to
${\rm trace} ( R^{1/2} K W_c K^\top R^{1/2} )$. 
As a natural counterpart of the continuous-time formulation
in \cite{feron1992numerical}, the optimal controller $K$
can be found by solving the optimization problem
\be\label{lqr-form}
\ba{l}
\min_{K,W,X}  \,\, {\rm trace}\left( Q_x W \right)  
+ {\rm trace} \left( X \right) \smallskip \\ 
\textrm{subject to} \smallskip \smallskip \\
\left\{
\def\arraystretch{1.3} 
\begin{array}{l}
(A+BK) W (A+BK)^\top  - W + I_n \preceq 0 \\ 
W \succeq I_n \\
X - R^{1/2} K W K^\top R^{1/2} \succeq 0
\end{array} \right.
\ea
\ee
This can be cast as a convex optimization problem by means of suitable change of variables \cite{feron1992numerical}.
Based on this formulation, it is straightforward to derive a data-dependent formulation of this
optimization problem. \smallskip 

\begin{theorem}\label{prop:data-driven-lqr}
Let condition \eqref{Ud} hold.
Then, the optimal $H_2$ state-feedback controller $K$ for system \eqref{eq:system_h2}
can be computed as 
$K = U_{0,1,T}Q (X_{0,T} Q )^{-1}$
where $Q$ optimizes
\be\label{lmi.lqr}
\ba{l}
\min_{Q,X}  \,\, {\rm trace}\left(Q_x X_{0,T} Q \right)+{\rm trace}\left(X \right) \smallskip
\\
\textrm{subject to} \smallskip  \\
\left\{
\def\arraystretch{1.3} 
\begin{array}{l}
\begin{bmatrix}
X &  R^{1/2} U_{0,1,T} Q \\
Q^\top U_{0,1,T}^\top R^{1/2} & X_{0,T} Q
\end{bmatrix}
\succeq 0 \smallskip \smallskip \\[4mm]
\begin{bmatrix}
X_{0,T} Q - I_n &  X_{1,T} Q \\
Q^\top X_{1,T}^\top & X_{0,T} Q
\end{bmatrix}
\succeq 0 
\end{array} %
\right.
\ea\ee
\end{theorem}
\smallskip 

{\em Proof.} 
In view of \eqref{compute.G_K} and 
the parametrization \eqref{data-control}, the optimal solution to \eqref{lqr-form} 
can be computed as $K = U_{0,1,T} G_K$, where
$G_K$ optimizes
\be\label{lqr-form2}
\ba{l}
\min_{G_K,W,X}  \,\, {\rm trace}\left( Q_x W \right)  
+ {\rm trace} \left( X \right) \smallskip \smallskip \\ 
\textrm{subject to} \smallskip \smallskip \\
\left\{
\def\arraystretch{1.3} 
\begin{array}{l}
X_{1,T} G_K W G_K^\top X_{1,T}^\top  - W + I_n \preceq 0 \\ 
W \succeq I_n \\
X - R^{1/2} U_{0,1,T} G_K W G_K^\top U_{0,1,T} G_K^\top R^{1/2} \succeq 0 \\
X_{0,T} G_K = I_n
\end{array} \right.
\ea
\ee
To see this, let $(K_*, W_*, X_*)$ be the optimal solution to \eqref{lqr-form}
with cost $J_*$. We show that the optimal solution $(\overline G_K,\overline W,\overline X)$ 
to \eqref{lqr-form2}
is such that $(K,W,X)=(U_{0,1,T} \overline G_K,\overline W,\overline X)$ is feasible 
for \eqref{lqr-form} and has cost $J_*$, which implies $K_* = U_{0,1,T} \overline G_K$ 
as the optimal controller is unique.
Feasibility simply follows from the fact that $K=U_{0,1,T} \overline G_K$
along with $X_{0,T} \overline G_K = I_n$ implies that $X_{1,T} \overline G_K = A+BK$.
In turn, this implies that $(K,W,X)=(U_{0,1,T} \overline G_K,\overline W,\overline X)$ satisfies all
the constraints in \eqref{lqr-form}. As a final step, let $\overline J$ be the cost associated with the 
solution $(K,W,X)=(U_{0,1,T} \overline G_K,\overline W,\overline X)$. Since the latter 
is a feasible solution to \eqref{lqr-form}, we must have
$\overline J \geq J_*$. Notice now that $\overline J$ is also the optimal cost 
of \eqref{lqr-form2} associated with the solution $(\overline G_K,\overline W,\overline X)$.
Accordingly, let $G_{K_*}$ be a solution to \eqref{compute.G_K} computed
with respect to $K=K_*$. Thus $(G_K,W,X) = (G_{K_*},W_*, X_*)$ is a feasible solution 
to \eqref{lqr-form2} with cost $J_*$. This implies that $\overline J \leq J_*$ and thus $\overline J = J_*$.
This shows that $K_* = U_{0,1,T} \overline G_K$.

The formulation \eqref{lmi.lqr} follows directly from \eqref{lqr-form2}
by defining $Q=G_K W$ and exploiting the relation
$X_{0,T} Q = W$.
\qedp \smallskip

 \emph{Illustrative example}. We consider the
batch reactor system of the previous subsection. 
As before,
we generate the data with random initial conditions and by applying to each input channel 
a random input sequence of length $T=15$ by using the MATLAB command \texttt{rand}.
We let $Q_x = I_n$ and $R = I_m$.
To solve \eqref{lmi.lqr}
we used CVX, obtaining 
\begin{eqnarray} 
K = \left[
\ba{rrrr}
0.0639 &  -0.7069  & -0.1572  & -0.6710 \\
    2.1481 &   0.0875  &  1.4899  & -0.9805
\ea
\right] \nonumber
\end{eqnarray}
This controller coincides with the controller $\overline K$ obtained
with the MATLAB command \texttt{dare} which solves the classic DARE equation. In particular,
$\|K-\overline K\| \approx 10^{-7}$. \qedp
\smallskip 
 
\begin{rem}
\emph{(Numerical issues for unstable systems)} 
The above results are implicitly based on open-loop data.
When dealing with unstable systems numerical instability problems 
may arise. Nonetheless, by Lemma \ref{lem:willems} 
a persistently exciting input of order $n+1$ suffices to ensure \eqref{Ud}. 
In turn (see the discussion in Section \ref{sec:db_reprsentation}), this ensures that
we ``only" need $T = (m+1) n +m$ samples in order to compute the controller. 
This guarantees that one 
can compute \emph{a priori} for how long a system should run in open loop.
In practice, this 
result also guarantees {practical applicability for systems of moderate size that are not strongly unstable}. 

When dealing with large scale and highly unstable systems the situation is inevitably more complex,
and other solutions might be needed. For instance, if a stabilising controller $\hat K$  
(not necessarily performing) is known, then one can think of running \emph{closed-loop} experiments  
during which a persistently exciting signal is superimposed  to the control signal given by $\hat K$, making sure that
all the previous results continue to follow without any modification. {Measures}
of this type are widely adopted in adaptive control  to overcome
issues of loss of stabilisability due to the lack of excitation caused by feedback \cite[Section 7.6]{Ioannou1996}. \qedp
\end{rem}

\section{
Robustness: noise-corrupted data
and nonlinear systems}
\label{sec:direct-data-driven_rob}

In the previous subsections, we have considered data-driven design
formulations based on LMIs. 
Besides their simplicity,  
one of the main reasons for resorting to
such formulations is that LMIs have proven their effectiveness 
also in the presence of perturbations and/or uncertainties 
around the system to be controlled \cite{scherer2004}. 
In this subsection, we exemplify this point by 
considering
stabilization with noisy data, as well as the problem of stabilizing an unstable equilibrium of 
a nonlinear system, which are both situations where identification can be challenging.  

\subsection{Stabilization with noisy data}

Consider again system \eqref{lin.sys.1}, but suppose
that one can only measure the signal
\begin{eqnarray}
\zeta(k) = x(k) + w(k) 
\end{eqnarray} 
where $w$ is an unknown measurement noise. 
We will assume no particular statistics on the noise. The problem of
interest is to design a stabilizing controller for system \eqref{lin.sys.1}
assuming that we measure $\zeta$.
Let
\begin{eqnarray}
&& W_{0, T} := \left[ \begin{array}{cccc} w_d(0) & w_d(1) & \cdots & w_d(T-1) \end{array} \right]\\
&& W_{1, T} := \left[ \begin{array}{cccc} w_d(1) & w_d(2) & \cdots & w_d(T) \end{array} \right]
\end{eqnarray} 
where $w_d(k)$, $k=0,1,\ldots,T$ are noise samples associated to the experiment, 
and
\begin{eqnarray}
&& Z_{0, T} := X_{0, T} + W_{0, T} \\
&& Z_{1, T} := X_{1, T} + W_{1, T}.
\end{eqnarray} 
The latter are the matrices containing the available information 
about the state of the system. 
Recall that in the noise-free 
case, a stabilizing controller can be found by searching for a solution $Q$ 
to the LMI \eqref{lmi.state.feedback}. In the noisy case,
it seems thus natural to replace \eqref{lmi.state.feedback} with the design condition
\begin{eqnarray} \label{noise_free_form}
\begin{bmatrix}
Z_{0,T} \,Q &  Z_{1,T} Q\\
Q^\top Z_{1,T}^\top & Z_{0,T} \,Q
\end{bmatrix}
\succ 0.
\end{eqnarray}
This condition already gives a possible solution approach. In fact, since
positive definiteness is preserved under sufficiently small perturbations, 
for every solution $Q$ to \eqref{lmi.state.feedback} there exists a noise 
level such that $Q$ will remain solution to \eqref{noise_free_form},
and such that the controller $K = U_{0,1,T}Q (Z_{0,T} Q )^{-1}$ obtained by replacing $X_{0,T}$
with $Z_{0,T}$ will remain stabilizing, where the latter property holds
since the eigenvalues of $A+BK$ depend with continuity on $K$.
This indicates that the considered LMI-based approach has some intrinsic degree of 
robustness to measurement noise.

We formalize these considerations by focusing the attention 
on a slightly different formulation, which 
consists in finding a matrix $Q$ and a scalar $\alpha > 0$
such that 
\begin{eqnarray} \label{lmi.state.feedback.noise}
\begin{bmatrix}
Z_{0,T} \,Q - \alpha Z_{1,T}  Z_{1,T}^\top &  Z_{1,T} Q\\
Q^\top Z_{1,T}^\top & Z_{0,T} \,Q
\end{bmatrix}
\succ 0, \quad \begin{bmatrix}
I_T & Q \\
Q^\top & Z_{0,T} \,Q
\end{bmatrix}
\succ 0 \nonumber \\
\end{eqnarray}
It is easy to verify that in the noise-free case 
and with persistently exciting inputs
also this formulation is always feasible
and any solution $Q$ is such that
$K = U_{0,1,T}Q (Z_{0,T} Q )^{-1}$
gives a stabilizing controller. We show this fact in the next Remark \ref{rem:feas_noise}.
We consider the formulation \eqref{lmi.state.feedback.noise} because it
makes it possible to explicitly \emph{quantify} noise levels 
for which a solution returns a stabilizing controller.} 
\smallskip

\begin{rem} \label{rem:feas_noise}
\emph{(Feasibility of \eqref{lmi.state.feedback.noise} under noise-free data)} 
In the noise-free case, that is when $Z_{0,T}=X_{0,T}$ and $Z_{1,T}=X_{1,T}$,
the formulations \eqref{noise_free_form} and \eqref{lmi.state.feedback} coincide.
Suppose then that \eqref{noise_free_form} is feasible and let $\overline Q$
be a solution. Since positive definiteness is preserved under 
small perturbations, $(Q,\alpha)=(\overline Q, \overline \beta)$ will be a solution to 
the first of \eqref{lmi.state.feedback.noise} for a sufficiently small $\overline \beta >0$.
Hence $(Q,\alpha)=(\delta \overline Q, \delta \overline \beta)$ will remain feasible 
for the first of \eqref{lmi.state.feedback.noise} for all $\delta >0$.
We can thus pick $\delta$ small enough so that 
$(Q,\alpha) :=(\delta \overline Q,\delta \overline \beta)$ satisfies
also the second of \eqref{lmi.state.feedback.noise}. 

Conversely, consider any
solution $(Q,\alpha)$ to \eqref{lmi.state.feedback.noise}
and let $K = U_{0,1,T}Q (Z_{0,T} Q )^{-1}$.
Since $\alpha>0$, the first inequality in \eqref{lmi.state.feedback.noise}
implies that \eqref{noise_free_form} also holds, which, in view of the identities
$Z_{0,T}=X_{0,T}$ and $Z_{1,T}=X_{1,T}$, is equivalent to have condition \eqref{lmi.state.feedback} satisfied.  Hence, the gain $K$ is stabilizing. 
\qedp
\end{rem}
\smallskip

Consider the following assumptions.

\smallskip
\begin{assumption} \label{ass:noise}
The matrices 
 \begin{eqnarray}
\begin{bmatrix}
U_{0,1,T} \\
\hline
Z_{0, T}
\end{bmatrix}, \;Z_{1, T}
\end{eqnarray} 
have full row rank. \qedp

\smallskip
\end{assumption}
\begin{assumption} \label{ass:noise2}
It holds that 
 \begin{eqnarray}
R_{0, T} R_{0, T}^\top \preceq \gamma Z_{1, T} Z_{1, T}^\top
\end{eqnarray} 
for some $\gamma > 0$, where $R_{0, T} := A W_{0, T} - W_{1, T}$. \qedp
\end{assumption} 
\smallskip

Assumptions  \ref{ass:noise} and  \ref{ass:noise2} 
both express the natural requirement that the 
loss of information caused by noise is not significant. 
In particular,
Assumption \ref{ass:noise} is the
counterpart of condition \eqref{Ud} for noise-free data,
and is always satisfied when the input is persistently exciting 
and the noise is sufficiently small. This is because:
(i) condition \eqref{Ud} implies that $X_{0, T}$ has 
rank $n$; (ii) $X_{1, T} = A X_{0, T} + B U_{0,1, T}$
so that condition \eqref{Ud} implies that ${\rm rank} \, X_{1, T} = {\rm rank} \left[ B \,\,\, A \right] = n$
otherwise the system would not be controllable; and
(iii) the rank of a matrix does not change under sufficiently small
perturbations.

Intuitively, Assumption \ref{ass:noise} alone is not sufficient to guarantee the existence of a solution 
returning a stabilizing controller since this assumption may
also be verified by 
arbitrary noise, 
in which case the data need not contain
any useful information. Assumption \ref{ass:noise2} takes into account 
this aspect, and plays the role of a ``\emph{signal-to-noise ratio}" (SNR) condition. 
Notice that when Assumption \ref{ass:noise} holds then 
Assumption \ref{ass:noise2} is always satisfied 
for large enough $\gamma$. As next theorem shows, however,
to get stability one needs to restrict the magnitude of $\gamma$,
meaning that the SNR must be sufficiently large.

\smallskip

\begin{theorem} \label{prop:noise_stability}
Suppose that  
Assumptions \ref{ass:noise} and \ref{ass:noise2} hold.
Then, any solution $(Q,\alpha)$ to \eqref{lmi.state.feedback.noise}
such that $\gamma < \alpha^2/(4 + 2\alpha)$ returns a 
stabilizing controller
$K = U_{0,1,T}Q (Z_{0,T} Q )^{-1}$.
\end{theorem} \smallskip

{\em Proof.} As a first step, 
we parametrize the closed-loop system as a function of 
$G_K$ and the noise,
{\setlength\arraycolsep{2pt} 
 \begin{eqnarray}
A+B K &=& 
\begin{bmatrix}
B & A
\end{bmatrix} 
\begin{bmatrix} K \\ I
\end{bmatrix} \nonumber \\
&=&
\begin{bmatrix}
B & A
\end{bmatrix} 
\begin{bmatrix}
U_{0,1,T} \\
\hline
Z_{0, T}
\end{bmatrix}
G_{ K} \nonumber \\
&=&
\begin{bmatrix}
B & A
\end{bmatrix} 
\begin{bmatrix}
U_{0,1,T} \\
\hline
X_{0, T} + W_{0, T}
\end{bmatrix}
G_{ K} \nonumber \\
&=&
X_{1, T} G_{ K} + A W_{0, T} G_{ K} \nonumber \\
&=&
\left( Z_{1, T} +  R_{0, T} \right) G_{ K} \label{48}
 \end{eqnarray}
where $G_{K}$ is a solution to
\be \label{compute.G_K.noise}
\begin{bmatrix}
 K \\ I
\end{bmatrix}
= 
\begin{bmatrix}
U_{0,1,T} \\
\hline
Z_{0, T}
\end{bmatrix}
G_{ K} 
\ee
which exists in view of Assumption \ref{ass:noise}.

By this parametrization, $A+BK$ is stable if and only if 
there exists $P\succ 0$ such that 
 \begin{eqnarray} \label{eq:prop.noise.Gk}
\left( Z_{1, T} +  R_{0, T} \right) G_{ K} P G_{ K}^\top \left( Z_{1, T} +  R_{0, T} \right)^\top - P \prec 0
 \end{eqnarray}
where $G_K$ satisfies \eqref{compute.G_K.noise}. 
Following the same analysis as in Section \ref{subsec:data-driven-parametrization},
introducing the change of variable $Q = G_{ K} P$ and exploiting 
the relation $Z_{0,T} Q = P$,
stability is equivalent 
to the existence of a matrix $Q$ such that 
{\setlength\arraycolsep{2.3pt} 
\begin{eqnarray} \label{eq:prop.noise.Q}
\def\arraystretch{1.3}
\left\{
\begin{array}{l}
{Z_{0,T} Q\succ 0}\\
 (Z_{1, T} +  R_{0, T}) Q (Z_{0,T} \,Q)^{-1} \cdot \\ 
 \qquad \cdot \, Q^\top (Z_{1, T} +  R_{0, T})^\top 
-  Z_{0,T} \,Q \prec 0  \\
 U_{0,1,T}Q = K Z_{0, T} Q 
\end{array} \right.
\end{eqnarray}}%
From the viewpoint of design, one can focus on the 
inequality {constraints},
since the equality constraint can be satisfied a posteriori with  
$K = U_{0,1,T}Q ({Z_{0,T}} Q )^{-1}$.

We can now finalize the proof. First
recall that for arbitrary matrices $X,Y,F$ with $F \succ 0$, and  a scalar $\varepsilon > 0$, 
it holds that $X F Y^\top + Y F X^\top \preceq \varepsilon X F X^\top + \varepsilon^{-1} Y F Y^\top$. 
By applying this property to the second inequality in \eqref{eq:prop.noise.Q}
with $F = Z_{0,T} Q$, $X = Z_{1, T} Q (Z_{0,T} Q)^{-1}$,
$Y = R_{0, T} Q (Z_{0,T} Q)^{-1}$,
a sufficient condition for stability is that 
{\setlength\arraycolsep{2.3pt}
\begin{eqnarray*} \label{eq:prop.noise.Q.1}
\def\arraystretch{1.3}
\left\{
\begin{array}{l}
{Z_{0,T} Q\succ 0}\\
\Theta :=
(1+\varepsilon) Z_{1, T} Q (Z_{0,T} \,Q)^{-1}  Q^\top Z_{1, T}^\top 
\\ 
\qquad +  (1+\varepsilon^{-1}) R_{0, T} Q (Z_{0,T} \,Q)^{-1}  Q^\top R_{0, T}^\top  -  Z_{0,T} \,Q \prec 0 
\end{array} \right.
\end{eqnarray*}}%
where $\varepsilon >0$.
By the Schur complement
any solution $(Q,\alpha)$ 
gives $Z_{1, T} Q (Z_{0,T} \,Q)^{-1}  Q^\top Z_{1, T}^\top +\alpha Z_{1, T} Z_{1, T}^\top -Z_{0,T} Q \prec 0$
and $Q (Z_{0,T} \,Q)^{-1}  Q^\top \prec I_T$.
Accordingly, any solution $(Q,\alpha)$ ensures that
\begin{equation} \label{eq:prop.noise.Q.2}
\Theta \prec
- \alpha Z_{1, T}  Z_{1, T}^\top  +
{\varepsilon} Z_{1, T} Z_{1, T}^\top + 
(1+\varepsilon^{-1}) R_{0, T} R_{0, T}^\top 
\end{equation}
This implies that any solution $(Q,\alpha)$ to \eqref{lmi.state.feedback.noise} ensures
stability if the right hand side of \eqref{eq:prop.noise.Q.2}
is negative definite.
Pick $\varepsilon = \alpha/2$.
The right hand side of \eqref{eq:prop.noise.Q.2}
is negative definite if
\[
R_{0, T} R_{0, T}^\top \prec \frac{\alpha^2}{2(2 + \alpha)}
Z_{1, T}  Z_{1, T}^\top 
\]
which is satisfied when $\gamma < \alpha^2/(4 + 2\alpha)$. \qedp \smallskip

\emph{Illustrative example}. We consider the
batch reactor system of the previous section.
We generate the data with unit random initial conditions and by applying to each input channel 
a unit random input sequence of length $T=15$. The 
noise is taken as a random sequence within $[-0.01,0.01]$.
To solve \eqref{lmi.state.feedback.noise} we used CVX, obtaining 
\begin{eqnarray} 
K = \left[
\ba{rrrr}
2.5934 &  -1.6853  &  3.2184 &  -1.8010\\
    3.1396 &   0.1146   & 3.2873  & -1.5069
\ea
\right] \nonumber
\end{eqnarray}
with $\alpha \approx 10^{-4}$. 
Condition $\gamma < \alpha^2/(4 + 2\alpha)$ is not satisfied 
as the smallest value of $\gamma$ satisfying Assumption \ref{ass:noise2}
is $\approx10^{-2}$. Nonetheless, 
$K$ stabilizes the closed-loop system. As pointed out, this 
simply reflects that the condition $\gamma < \alpha^2/(4 + 2\alpha)$ 
can be theoretically conservative. 
In fact, numerical simulations indicate that
condition $\gamma < \alpha^2/(4 + 2\alpha)$ is satisfied for noise of order $10^{-4}$,
while in practice the algorithm systematically returns stabilizing controllers
for noise of order $10^{-2}$, and for noise of order $10^{-1}$ 
(noise which can also alter the first digit of the noise-free trajectory) 
it returns stabilizing controllers in more than half of the cases.
\qedp 

\smallskip

In contrast with Assumption \ref{ass:noise} which can be
assessed from data only, checking whether Assumption \ref{ass:noise2} holds with
a value $\gamma < \alpha^2/(4 + 2\alpha)$
requires prior knowledge of an upper bound on $R_{0,T}$.
In turn, this requires prior knowledge of an upper bound on the noise and on
the largest singular value of $A$.
If this information is available then Assumption \ref{ass:noise2}
can be assessed from data.
\footnote{
For instance, 
recalling that $W_{0,T}$ 
and $W_{1,T}$ are $n \times T$ matrices, it follows 
from the Gershgorin theorem that
\begin{eqnarray}
W_{0,T} W_{0,T}^\top \preceq n \overline w T I_n, \quad
W_{1,T} W_{1,T}^\top \preceq n \overline w T I_n
\end{eqnarray}
where 
$\overline w$ denotes an upper bound on the noise, that is
$|w_i(k) w_j(k)|\le \overline w$ for all $1\le i,j\le n$ and for all $k=0,1,\ldots, T$.
This implies that $R_{0,T}$ satisfies
$R_{0,T} R_{0,T}^\top \preceq  2 n \overline w T I_n (1+\sigma_A)$,
where $\sigma_A$ denotes the 
the square of the largest singular value of the matrix $A$.}
One can replace Assumption \ref{ass:noise2} with a (more conservative) condition which
can be assessed under the only assumption that an upper bound on the noise
is available.
Before stating this result, we nonetheless point out that there is a 
reason why $A$ appears in Assumption \ref{ass:noise2}.
In fact, the information loss caused by noise
does not depend only on the magnitude of the noise but also on its ``direction".
For instance, in case the noise $w$ follows the equation $w(k+1) = A w(k)$ then
$R_{0, T}$ becomes zero, meaning that Assumption \ref{ass:noise2} holds with 
an arbitrary $\gamma$
irrespective of the magnitude 
of $w$. In fact, in this case $w$ behaves as a genuine system trajectory 
(it evolves in the set of states that the system can generate),
so it brings useful information on the system dynamics. 
This indicates that noise of large magnitude but ``close"
to the set of states where the system evolves can be less detrimental of noise with smaller magnitude
but which completely alters the direction of the noise-free trajectory. 
 
As anticipated, one can replace Assumption \ref{ass:noise2} 
with a (more conservative) condition verifiable
under the only assumption that an upper bound on the noise
is known.

\smallskip
\begin{assumption} \label{ass:noise3}
It holds that 
\begin{eqnarray} \label{ass:noise3_a}
&& \begin{bmatrix}
0 \\ \hline W_{0, T}
\end{bmatrix}
\begin{bmatrix}
0 \\ \hline W_{0, T}
\end{bmatrix}^\top  \preceq \gamma_1 
\begin{bmatrix}
U_{0,1,T} \\
\hline
Z_{0, T}
\end{bmatrix}
\begin{bmatrix}
U_{0,1,T} \\
\hline
Z_{0, T}
\end{bmatrix}^\top \\[0.2cm]
\label{ass:noise3_b}
&& W_{1, T} W_{1, T}^\top \preceq \gamma_2 Z_{1, T} Z_{1, T}^\top
\end{eqnarray} 
for some $\gamma_1 \in (0,0.5)$ and $\gamma_2 > 0$.
\qedp
\end{assumption} 
\smallskip

\smallskip
\begin{corollary} \label{prop:noise_stability_2} 
Suppose that  
Assumptions \ref{ass:noise} and \ref{ass:noise3} hold.
Then, any solution $(Q,\alpha)$ to \eqref{lmi.state.feedback.noise}
such that  
\begin{eqnarray}
\frac{6 \gamma_1+3\gamma_2}{1 - 2 \gamma_1}
<
\frac{\alpha^2}{2(2 + \alpha)}
\end{eqnarray}
returns a 
stabilizing controller
$K = U_{0,1,T}Q (Z_{0,T} Q )^{-1}$. 
\end{corollary} 
\smallskip

\emph{Proof.} 
See the Appendix. \qedp

\smallskip
In both Theorem \ref{prop:noise_stability} and Corollary \ref{prop:noise_stability_2}, 
stability relies on the fulfilment 
of a condition like $\gamma < \alpha^2/(4 + 2\alpha)$. 
This suggests that it might be convenient to reformulate the 
design problem by searching for the solution $(Q,\alpha)$ to \eqref{lmi.state.feedback.noise} 
maximizing $\alpha$,
which still results in a convex problem. Nonetheless, it is worth noting that both
Theorem \ref{prop:noise_stability} and Corollary \ref{prop:noise_stability_2} only
give \emph{sufficient} conditions,
meaning (as shown also in the previous numerical example)
that one can find stabilizing controllers even when
$\gamma \geq \alpha^2/(4 + 2\alpha)$.

\subsection{Stabilization of nonlinear systems}

The previous result shows that a controller can be designed 
in the presence of noise provided that signal-to-noise ratio 
is sufficiently small. This hints at the 
possibility of  designing also a stabilizing control for  nonlinear 
systems based on data alone. As a matter of fact, around an 
equilibrium a nonlinear system can be expressed via its first 
order approximation plus a reminder. If we run our experiment 
in such a way that the input and the state remain sufficiently 
close to the equilibrium, then the reminder can be viewed as 
a process disturbance of small magnitude and 
there is a legitimate hope that the robust stabilization result 
also applies to this case. 
In the rest of this section we formalize this intuition. 

Consider a smooth nonlinear system 
\be\label{nonl}
x(k+1) = f(x(k),u(k))
\ee
and let $(\overline x, \overline u)$ be a \emph{known} equilibrium pair, 
that is such that 
$\overline x = f(\overline x, \overline u)$.
{Let us rewrite the nonlinear system as}
\begin{eqnarray} \label{nonl2}
\delta x(k+1) = A \delta x(k)  + B \delta u(k) + d(k)
\end{eqnarray}
where $\delta x := x- \overline x$, $\delta u := u-\overline u$, and where
\begin{eqnarray} \label{lin_matrices}
A:=\left.\frac{\partial f}{\partial x}\right|_{(x,u)=(\overline x, \overline u)},\quad
B:=\left.\frac{\partial f}{\partial u}\right|_{(x,u)=(\overline x, \overline u)} \,.
\end{eqnarray}
The quantity $d$ accounts for higher-order terms and it has the property that is 
goes to zero faster than $\delta x$ and $\delta u$, namely we have
\begin{eqnarray*} 
d=R(\delta x, \delta u)\begin{bmatrix} \delta x \\ \delta u\end{bmatrix}
\end{eqnarray*}
with $R(\delta x, \delta u)$ an $n\times (n+m)$ matrix of smooth functions with the property that 
\begin{eqnarray} \label{eq:convergence_d}
\lim_{{\tiny \begin{bmatrix} \delta x \\ \delta u\end{bmatrix}\to {0}}} R(\delta x, \delta u)= {0}
\end{eqnarray}

It is known that if the pair $(A,B)$ defining the linearized 
system is stabilizable then the controller $K$ rendering $A+BK$ 
stable also exponentially stabilizes the equilibrium $(\overline x, \overline u)$  for the original nonlinear system. 
The objective here is to provide sufficient 
conditions for the design of $K$ from data. To this end, we consider the following result
which is an adaptation of Theorem \ref{prop:noise_stability}. Let  
\begin{eqnarray*}
X_{0, T} &:=& \left[ \begin{array}{cccc} \delta x_d(0) & \delta x_d(1) & \cdots & \delta x_d(T-1) \end{array} \right] \\
X_{1, T} &:=& \left[ \begin{array}{cccc} \delta x_d(1) & \delta x_d(2) & \cdots & \delta x_d(T) \end{array} \right]  \\
U_{0,1,T} &:=& \left[ \begin{array}{cccc} \delta u_d(0) & \delta u_d(1) & \cdots & \delta u_d(T-1) \end{array} \right] \\
D_{0, T} &:=& \left[ \begin{array}{cccc} d_d(0) & d_d(1) & \cdots & d_d(T-1) \end{array} \right]
\end{eqnarray*} 
be the data resulting from an experiment carried out on the nonlinear system \eqref{nonl}.
Note that the matrices $X_{0, T}$, $X_{1, T}$ and 
$U_{0,1,T}$ are known. Consider the following assumptions.

\smallskip 
\begin{assumption} \label{ass:dist}
The matrices 
 \begin{eqnarray}
\begin{bmatrix}
U_{0,1,T} \\
\hline
X_{0, T}
\end{bmatrix}, \;X_{1, T}
\end{eqnarray} 
have full row rank. \qedp
\smallskip

\end{assumption}
\begin{assumption} \label{ass:dist2}
It holds that 
 \begin{eqnarray}
D_{0, T} D_{0, T}^\top \preceq \gamma X_{1, T} X_{1, T}^\top
\end{eqnarray} 
for some $\gamma >0$. \qedp
\end{assumption} 
\smallskip

The following result holds.  

\smallskip
\begin{theorem} \label{prop:nonlinear_stability} 
Consider a nonlinear system as in \eqref{nonl}, 
along with an equilibrium pair $(\overline x, \overline u)$.
Suppose that  
Assumptions \ref{ass:dist} and \ref{ass:dist2} hold.
Then, any solution $(Q,\alpha)$ to
\begin{eqnarray} \label{lmi.state.feedback.nonlinear}
\begin{bmatrix}
X_{0,T} \,Q - \alpha X_{1,T}  X_{1,T}^\top &  X_{1,T} Q\\
Q^\top X_{1,T}^\top & X_{0,T} \,Q
\end{bmatrix}
\succ 0, \quad \begin{bmatrix}
I_T & Q \\
Q^\top & X_{0,T} \,Q
\end{bmatrix}
\succ 0 \nonumber \\
\end{eqnarray}
such that $\gamma < \alpha^2/(4 + 2\alpha)$ returns a 
stabilizing state-feedback gain
$K = U_{0,1,T}Q (X_{0,T} Q )^{-1}$, which locally stabilizes 
the equilibrium pair $(\overline x, \overline u)$.
\end{theorem}

\smallskip

\emph{Proof.} We only sketch the proof since essentially analogous 
to the proof of Theorem \ref{prop:noise_stability}.
Note that 
{\setlength\arraycolsep{2pt} 
 \begin{eqnarray}
A+B K &=& 
\begin{bmatrix}
B & A
\end{bmatrix} 
\begin{bmatrix} K \\ I
\end{bmatrix} =
\begin{bmatrix}
B & A
\end{bmatrix} 
\begin{bmatrix}
U_{0,1,T} \\
\hline
X_{0, T}
\end{bmatrix}
G_{ K} \nonumber \\
&=&
\left( X_{1, T} -  D_{0, T} \right) G_{ K} \label{48}
 \end{eqnarray}}
where $G_{K}$ is a solution to
\be \label{compute.G_K.nonlinear}
\begin{bmatrix}
 K \\ I
\end{bmatrix}
= 
\begin{bmatrix}
U_{0,1,T} \\
\hline
X_{0, T}
\end{bmatrix}
G_{ K} 
\ee
which exists in view of Assumption \ref{ass:dist}. The rest of 
the proof follows exactly the same steps as the proof of Theorem \ref{prop:noise_stability} 
by replacing $Z_{0, T}$, $Z_{1, T}$ and $R_{0, T}$ by $X_{0, T}$, $X_{1, T}$ and $-D_{0, T}$,
respectively. \qedp 

\smallskip

Before illustrating the result with a numerical example, we make some 
observations.

Assumptions \ref{ass:dist} and \ref{ass:dist2} parallel the assumptions 
considered for the case of noisy data.
In particular, Assumptions \ref{ass:dist2} is the counterpart 
of Assumption \ref{ass:noise2} (or Assumption \ref{ass:noise3})
and it amounts to requiring that the experiment is carried out sufficiently close to
the system equilibrium so that the effect of the nonlinearities 
(namely the disturbance $d$) becomes small enough compared with $\delta x$
(\emph{cf.} \eqref{eq:convergence_d}). 

At this moment, we do not have a method for designing the experiments 
in such a way that Assumptions \ref{ass:dist} and \ref{ass:dist2} hold.
This means that verifying Assumption \ref{ass:dist2} requires at this stage prior knowledge of an upper bound 
on $d$, that is on the type of nonlinearity
(Assumption \ref{ass:dist} can be anyway assessed from data only). 
Albeit in some cases this information can
be inferred from physical considerations, in general 
this is an important aspect which deserves to be studied.
Numerical simulations (including the example which follows) 
nonetheless indicate that at least in certain cases the ``margin" is appreciable in 
the sense that one obtains stabilizing controllers even when the experiment leads
the system sensibly far from its equilibrium. 
\smallskip

\emph{Illustrative example.} 
Consider the Euler discretization of an inverted pendulum
\[\ba{rl}
x_1(k+1) =& x_1(k)+ \Delta x_2(k)\\
x_2(k+1) =& \displaystyle  \frac{\Delta  g}{\ell} \sin x_1(k) + \left( 1 - \frac{\Delta  \mu}{m \ell^2} \right) x_2(k) 
+\frac{\Delta }{m \ell^2}u(k)\\
\ea\]
where we simplified the sampled times $k \Delta$ in $k$, with $\Delta$ the sampling time. The states $x_1, x_2$ are the angular position and velocity, respectively, $u$ is the applied torque.
The system has an unstable equilibrium in 
$(\overline x, \overline u)=(0,0)$ corresponding to the pendulum upright position
and therefore $\delta x= x$ and $\delta u=u$.
It is straightforward to verify that 
\[
d(k)=
\begin{bmatrix}
0 \\
\displaystyle \frac{\Delta g}{\ell} \left(\sin x_1(k) - x_1(k) \right)
\end{bmatrix}
\]

Suppose that the parameters are 
$\Delta= 0.1$, 
$m=\ell=1$, $g=9.8$ and $\mu=0.01$.
The control design procedure is implemented in 
MATLAB. We generate the data with random initial conditions 
within $[-0.1,0.1]$,
and by applying a random input sequence of length $T=5$ within $[-0.1,0.1]$.
To solve \eqref{lmi.state.feedback.nonlinear} we used CVX, obtaining 
\begin{eqnarray} 
K = \left[
\ba{rr}
-12.3895  &  -3.6495 
\ea
\right], \nonumber
\end{eqnarray}
which stabilizes the unstable equilibrium in agreement with
Theorem \ref{prop:nonlinear_stability} as 
the linearized system has matrices
\[
A = 
\begin{bmatrix}
1.0000 &   0.1000 \\
    0.9800  &  0.9990
\end{bmatrix}, \quad 
B = \begin{bmatrix}
0 \\ 0.1
\end{bmatrix}
\]
In this example, $\alpha = 0.0422$ and
condition $\gamma < \alpha^2/(4 + 2\alpha)$ holds
because $X_{1,T}$ is of order $0.01$ and
$D_{0,T}$ is of order $10^{-5}$ so that
the smallest value of $\gamma$ for which {Assumption \ref{ass:dist2}} holds
is $\approx10^{-6}$ while $\alpha^2/(4 + 2\alpha) \approx10^{-4}$. 
We finally notice that the algorithm systematically returns stabilizing controllers
also for initial conditions and inputs within the interval $[-0.5,0.5]$ which corresponds
to an initial displacement of about $28$ degrees from the equilibrium, albeit in this case 
condition $\gamma < \alpha^2/(4 + 2\alpha)$ not always holds.
\qedp

\section{Input-output data: the case of SISO systems}\label{sec:of-siso}

In Section \ref{subsec:data-driven-parametrization}, 
the measured data are the inputs and the state, and the starting point is to express 
the trajectories of the system and the control gain in terms of the Hankel matrix of input-state data.  
Here we show how similar arguments can be used when only 
input/output data are accessible. The main derivations
 are given for single-input single-output (SISO) systems. A remark on
multi-input multi-output (MIMO) systems is provided in Section \ref{sec:MIMO}.

Consider a SISO systems as in \eqref{lin.sys} in left difference operator representation 
\cite[Section 2.3.3]{goodwin2014},
\be\label{eq:difference}
\ba{r}
y(k)+a_{n} y(k-1)+\ldots + a_{2} y(k-n+1)+ a_{1} y(k-n) \smallskip \\
= 
b_{n} u(k-1)+\ldots + b_{2} u(k-n+1)+ b_{1} u(k-n)
\ea\ee
This representation corresponds to \eqref{lin.sys} for $D=0$. 
In this case, one can 
reduce the output measurement case to the state measurement case with minor effort.
Let
\be \label{chi}
\ba{r}
\chi (k) := {\rm col} 
(y(k-n), y(k-n+1), \ldots, y(k-1), \smallskip \\
u(k-n), u(k-n+1), \ldots, u(k-1)),
\ea\ee
from \eqref{eq:difference} 
we obtain the  state space system \eqref{eq:new-state-space-sys} on the next page.
Note that we turned our attention to a system of order $2n$, which is not minimal. 
\begin{figure*}[!t]
\normalsize
{
\setcounter{MaxMatrixCols}{20}
\be\label{eq:new-state-space-sys}\ba{rl}
\chi(k+1) 
= &
\underbrace{
\begin{bmatrix}
0 & 1 & 0 & \cdots &  0
&
0 & 0  & 0
& \cdots  & 0
\\
0 & 0 & 1 & \cdots & 0
&
0 & 0 & 0 & \cdots  & 0
\\
\vdots & \vdots &  \vdots  & \ddots &  \vdots 
&
\vdots & \vdots &  \vdots  & \ddots  & \vdots 
\\
0  & 0  & 0  & \cdots &  1 
&
0 & 0  & 0 & \cdots  & 0
\\
-a_1 &-a_2 & -a_3 & \cdots & -a_{n} &
b_1 &b_2 & b_3 & \cdots  & b_{n} 
\\
\hline
0  & 0   & 0  & \cdots & 0  
&
0 & 1  & 0 & \cdots  & 0
\\
0  & 0   & 0  & \cdots &  0  
&
0 &  0   & 1
& \cdots 
& 0
\\
\vdots & \vdots &  \vdots  & \ddots & \vdots 
&
\vdots & \vdots &  \vdots  & \ddots & \vdots 
\\
0  & 0   & 0  & \cdots & 0  
&
0  &  0    & 0 
& \cdots   
& 1
\\
0  & 0   & 0  & \cdots &  0  
&
0  &  0    & 0 
& \cdots
& 0 
\end{bmatrix}
}_{\mathcal{A}}
\chi(k)+
\underbrace{
\begin{bmatrix}
0
\\
0
\\
\vdots 
\\
0
\\
0
\\
\hline
0  
\\
0  
\\
\vdots 
\\
0  
\\
1\end{bmatrix}
}
_{\mathcal{B}}
u(k)
\\[2cm] 
y(k) = & 
\underbrace{
\begin{bmatrix}
-a_1 &-a_2 & -a_3 & \cdots & -a_{n} &
b_1 &b_2 & b_3 & \cdots & b_{n} 
\end{bmatrix}
}_{\mathcal{C}}
\chi(k)
\ea
\ee}
\hrulefill
\vspace*{4pt}
\end{figure*}
Consider now the matrix in {\eqref{Ud}
written for the system $\chi(k+1)=\mathcal{A}\chi(k)+\mathcal{B} u(k)$ 
in \eqref{eq:new-state-space-sys}, with $T$ satisfying $T\ge 2n+1$.
If this matrix is  full-row rank, then the analysis 
in the previous sections can be repeated also for system \eqref{eq:new-state-space-sys}. 
For system \eqref{eq:new-state-space-sys}
the matrix in question takes the form
\be\label{matrix.in.question}
\begin{bmatrix}
U_{0,1,T} \\
\hline
\hat X_{0, T}
\end{bmatrix}
= 
\begin{bmatrix}
u_d(0) & u_d(1) & \ldots & u_d(T-1)\\
\hline
\chi_{d}(0) & \chi_{d}(1)  & \ldots & \chi_{d}(T-1)\\
\end{bmatrix},
\ee
where $\chi_{d}(i+1)=\mathcal{A} \chi_{d}(i)+\mathcal{B} u_{d}(i)$
for $i\geq0$ and where $\chi_{d}(0)$ is the initial condition in the experiment, 
\[
\ba{r}
\chi_{d}(0) =
{\rm col}
(y_d(-n),
y_d(-n+1),
\ldots,
y_d(-1),
\smallskip \\
u_d(-n),
u_d(-n+1),
\ldots,
u_d(-1)
).
\ea
\]
The following result holds. \smallskip

\begin{lem}\label{lem:lin.independence.data}
The identity 
\be\label{aux0}
\begin{bmatrix}
U_{0,1,T} \\
\hline
\hat X_{0, T}
\end{bmatrix}
=
\begin{bmatrix}
U_{0,1,T}\\
\hline
Y_{-n,n,T}\\
\hline
U_{-n,n,T}\\
\end{bmatrix}
\ee
holds. 
Moreover, 
if {$u_{d, [0, T-1]}$} is persistently exciting of order $2n+1$ then 
\be\label{rank.cond.of}
{\rm rank}
\begin{bmatrix}
U_{0,1,T} \\
\hline
\hat X_{0, T}
\end{bmatrix}
=
2n+1.
\ee
\end{lem} \smallskip

{\em Proof.} 
The identity \eqref{aux0} follows immediately from the definition of the state $\chi$ in \eqref{chi} and the 
definition $\hat X_{0,T}$ in \eqref{matrix.in.question}. As for the second statement,
by the Key Reachability Lemma \cite[Lemma 3.4.7]{goodwin2014}, it is known that the 
$2n$-dimensional state space model \eqref{eq:new-state-space-sys} is controllable if and only if the 
polynomials 
$z^n+ a_n z^{n-1}\ldots+ a_2 z + a_1$, $b_n z^{n-1}+ \ldots+ b_2 z + b_1$ are coprime. 
Under this condition and persistency of excitation, Lemma \ref{lem:willems} applied to 
\eqref{eq:new-state-space-sys} immediately proves \eqref{rank.cond.of}. \qedp

\subsection{Data-based open-loop representation}

Similar to the case in which inputs and states are measured, the full rank property 
\eqref{rank.cond.of} plays a crucial role in expressing the system via data. As a matter of fact, 
for any pair $(u,\chi)$ we have 
\begin{eqnarray}
\label{aux1-output}
\begin{bmatrix} u \\ \chi\end{bmatrix}
=
\begin{bmatrix}
U_{0,1,T} \\
\hline
\hat X_{0, T}
\end{bmatrix}
g
\end{eqnarray}
for some $g$.
Hence,
\begin{eqnarray}
\label{important3-output}
\chi(k+1) &=&
\begin{bmatrix}
\mathcal{B} & \mathcal{A}
\end{bmatrix}
\begin{bmatrix}
u(k)\\
\chi(k)
\end{bmatrix}
\nonumber \\
&=& 
\begin{bmatrix}
\mathcal{B} & \mathcal{A}
\end{bmatrix}
\begin{bmatrix}
U_{0,1,T} \\
\hline
\hat X_{0, T}
\end{bmatrix}
g
= 
\hat X_{1, T} g
\end{eqnarray}
where 
\begin{eqnarray}
\label{hat.X}
\hat X_{1,T}  =
\begin{bmatrix}
Y_{-n+1,n,T}\\
\hline
U_{-n+1,n,T}\\
\end{bmatrix},
\quad
\hat X_{0,T}=
\begin{bmatrix}
Y_{-n,n,T}\\
\hline
U_{-n,n,T}\\
\end{bmatrix}.
\end{eqnarray}
As in the proof of Theorem \ref{prop:sys.identif} for the full state measurement case, 
we can thus solve for $g$ in \eqref{aux1-output}, replace it in \eqref{important3-output}, 
and obtain the following result.

\smallskip
\begin{theorem}
Let condition \eqref{rank.cond.of} hold.
Then system
\eqref{eq:new-state-space-sys}
has the following equivalent representation:
\be\label{data-dependent-model}
\ba{rl}
\chi(k+1) = &\hat X_{1,T}  \begin{bmatrix}
U_{0,1,T}\\
\hline
\hat X_{0,T}
\end{bmatrix}
^\dag
\begin{bmatrix}
u(k)\\
\chi(k)
\end{bmatrix} \smallskip \\
y(k) = &
e_{n}^\top
\hat X_{1,T} 
\begin{bmatrix}
U_{0,1,T}\\
\hline
\hat X_{0,T}
\end{bmatrix}
^\dag
\begin{bmatrix}
0_{1\times 2n}\\
I_{2n}
\end{bmatrix}
\chi(k)
\ea\ee
with $e_n$ the $n$-th versor of $\mathbb{R}^{2n}$.
\end{theorem} \smallskip

{\em Proof.} The proof follows the same steps as the proof of 
Theorem \ref{prop:sys.identif} and is omitted. \qedp
\smallskip

A representation of order $n$ of the system can also be extracted 
from \eqref{data-dependent-model}.
The model \eqref{data-dependent-model}, which only depends 
on measured input-output data, can be used for various analysis 
and design purposes. In the next subsection, 
we focus on the problem of designing an output feedback controller 
without going through the step of identifying a parametric model of the system. 

\subsection{Design of output feedback controllers}\label{subsec:dd-of}

Consider the left difference operator representation \eqref{eq:difference}, its realization 
\eqref{eq:new-state-space-sys} and the input/state pair $(u,\chi)$. 
We introduce a controller of the form
\be\label{eq:difference-control-}
\ba{r}
y^c(k)+c_{n} y^c(k-1)+\ldots + c_{2} y^c(k-n+1)+ c_{1} y^c(k-n) \smallskip \\
=
d_{n} u^c(k-1)+\ldots + d_{2} u^c(k-n+1)+ d_{1} u^c(k-n) \smallskip
\ea\ee
whose state space representation is given by 
\eqref{eq:new-state-space-sys-cont},
\begin{figure*}[!t]
\normalsize
{
\setcounter{MaxMatrixCols}{20}
\be\label{eq:new-state-space-sys-cont}\ba{rl}
\chi^c(k+1) 
= &
\underbrace{
\begin{bmatrix}
0 & 1 & 0 & \cdots & 0
&
0 & 0  & 0
& \cdots 
& 0
\\
0 & 0 & 1 & \cdots & 0
&
0 & 0 & 0 & \ldots & 0
\\
\vdots & \vdots &  \vdots  & \ddots & \vdots 
&
\vdots & \vdots &  \vdots  & \ddots & \vdots 
\\
0  & 0  & 0  & \cdots & 1 
&
0 & 0  & 0 & \cdots & 0
\\
-c_1 &-c_2 & -c_3 & \cdots  &-c_{n} &
d_1 &d_2 & d_3 & \cdots  & d_{n} 
\\
\hline
0  & 0   & 0  & \cdots  & 0  
&
0 & 1  & 0
& \cdots 
& 0
\\
0  & 0   & 0  & \cdots  & 0  
&
0 &  0   & 1
& \cdots 
& 0
\\
\vdots & \vdots &  \vdots  & \ddots &  \vdots 
&
\vdots & \vdots &  \vdots  & \ddots & \vdots 
\\
0  & 0   & 0  & \cdots   & 0  
&
0  &  0    & 0 
& \cdots 
& 1
\\
0  & 0   & 0  & \cdots  & 0  
&
0  &  0    & 0 
& \cdots   
& 0 
\end{bmatrix}
}_{\mathcal{F}}
\chi^c(k)+
\underbrace{
\begin{bmatrix}
0
\\
0
\\
\vdots 
\\
0
\\
0
\\
\hline
0  
\\
0  
\\
\vdots 
\\
0  
\\
1\end{bmatrix}
}
_{\mathcal{G}}
u^c(k)
\\[2cm]
y^c(k) = & 
\underbrace{
\begin{bmatrix}
-c_1 &-c_2 & -c_3 & \cdots & -c_{n} &
d_1 &d_2 & d_3 & \cdots &  d_{n} 
\end{bmatrix}
}_{\mathcal{H}}
\chi^c(k),
\ea
\ee 
}
\hrulefill
\vspace*{4pt}
\end{figure*}
with state $\chi^c$ defined similar to \eqref{chi}. 
In the closed-loop system, we enforce the following interconnection 
conditions relating the process and the controller
\be\label{interc.cond}
u^c(k)= y(k)\quad y^c(k)=u(k), \quad \; k\ge 0.
\ee
Note in particular the identity, for $k\ge n$, 
\be\label{eq:chi=controller.state}
\chi(k)=
\begin{bmatrix}
y_{[k-n, k-1]}\\
u_{[k-n, k-1]}
\end{bmatrix}
=
\begin{bmatrix}
u^c_{[k-n, k-1]}\\
y^c_{[k-n, k-1]}
\end{bmatrix}
=
\begin{bmatrix}
\mathbb{0}_{n\times n} & I_n\\
I_n & \mathbb{0}_{n\times n}
\end{bmatrix}
\chi^c(k). 
\ee
Hence, for $k\ge n$, there is no loss of generality in considering as the closed-loop system the system 

\smallskip
{\footnotesize 
\be\label{eq:new-state-space-sys-closed-loop}\ba{rl}
\chi(k+1) 
= &
\begin{bmatrix}
0 & 1 & 0 & \cdots & 0 
&
0 & 0  & 0
& \cdots 
& 0
\\
0 & 0 & 1 & \cdots & 0 
&
0 & 0 & 0 & \cdots & 0
\\
\vdots & \vdots &  \vdots  & \ddots &  \vdots 
&
\vdots & \vdots &  \vdots  & \ddots & \vdots 
\\
0  & 0  & 0  & \cdots &  1 
&
0 & 0  & 0 & \cdots & 0 
\\
-a_1 &-a_2 & -a_3 & \cdots & -a_{n} &
b_1 &b_2 & b_3 & \cdots & b_{n} 
\\
\hline
0  & 0   & 0  & \cdots   & 0  
&
0 & 1  & 0
& \cdots 
& 0
\\
0  & 0   & 0  & \cdots   & 0  
&
0 &  0   & 1
& \cdots 
& 0
\\
\vdots & \vdots &  \vdots  & \ddots &  \vdots 
&
\vdots & \vdots &  \vdots  & \ddots & \vdots 
\\
0  & 0   & 0  & \cdots & 0    
&
0 &  0   & 0
& \cdots 
& 1
\\
d_1 & d_2 & d_3 & \ldots &  d_{n} &
-c_1 &-c_2 & -c_3 & \ldots & -c_{n}
\end{bmatrix}
\chi(k).
\ea
\ee
}%

In the following result we say that controller  \eqref{eq:difference-control-} 
stabilizes system \eqref{eq:difference}, meaning that the closed-loop system 
\eqref{eq:new-state-space-sys-closed-loop} is asymptotically stable. 

\begin{theorem}\label{output-feedback-controller}
Let condition \eqref{rank.cond.of} hold. Then
the following properties hold:
\begin{itemize}
\item[(i)] The closed-loop system \eqref{eq:new-state-space-sys-closed-loop} has the  equivalent representation 
\be\label{cl.dynamic}
\chi(k+1) = \hat X_{1, T}
G_{\mathcal{K}}
\chi(k),
\ee
where $G_{\mathcal{K}}$ is a 
$T\times 2n$ matrix such that 
\be\label{non.so}
\begin{bmatrix}
\mathcal{K}\\
I_{2n}
\end{bmatrix}
=
\begin{bmatrix}
U_{0,1,T} \\
\hline
\hat X_{0, T}
\end{bmatrix}
G_{\mathcal{K}},
\ee
and 
\be\label{cal.K}
\mathcal{K}:=
\begin{bmatrix}
d_{1} & \ldots & d_{n} 
-c_{1} & \ldots & -c_{n} 
\end{bmatrix}
\ee
is the vector of coefficients of 
the controller \eqref{eq:difference-control-}.
\item[(ii)] Any matrix $\mathcal{Q}$ satisfying 
\begin{eqnarray}
\label{lmi.output.feedback}
\begin{bmatrix}
\hat X_{0,T} \,\mathcal{Q} &  \hat X_{1,T} \mathcal{Q}\\
\mathcal{Q}^\top \hat X_{1,T}^\top & \hat X_{0,T} \,\mathcal{Q}
\end{bmatrix}
\succ 0,
\end{eqnarray}
is such that the controller \eqref{eq:difference-control-} with coefficients given by 
\begin{eqnarray}
\label{data-control-explicit-of}
\mathcal{K} = U_{0,1,T} \mathcal{Q} (\hat X_{0,T} \mathcal{Q} )^{-1}
\end{eqnarray}
stabilizes system \eqref{eq:difference}. 
Conversely, any controller \eqref{eq:difference-control-} 
that stabilizes system \eqref{eq:difference} must have coefficients $\mathcal{K}$ 
given by \eqref{data-control-explicit-of}, with $\mathcal{Q}$ a solution of
\eqref{lmi.output.feedback}.
\end{itemize}
\end{theorem} \smallskip

{\em Proof.} (i) 
In view of condition \eqref{rank.cond.of} and by
Rouch\'{e}-Capelli theorem, a $T\times 2n$ matrix 
$G_{\mathcal{K}}$ exists such that \eqref{non.so} holds. 
Hence, 
{\setlength\arraycolsep{2.3pt}
\begin{eqnarray} 
\label{new.form.3}
\ba{rl}
\mathcal A+ \mathcal B \mathcal K 
=&
\begin{bmatrix}
\mathcal B & \mathcal A
\end{bmatrix}
\begin{bmatrix}
\mathcal K \\ I_{2n}
\end{bmatrix}
\\
=& \begin{bmatrix}
\mathcal B & \mathcal A
\end{bmatrix}
\begin{bmatrix}
U_{0,1,T} \\
\hline
\hat X_{0, T}
\end{bmatrix}
G_{\mathcal{K}} 
\\
= & 
\hat X_{1, T}
G_{\mathcal{K}} 
\ea
\end{eqnarray}}%
from which we obtain \eqref{cl.dynamic},
which are the dynamics \eqref{eq:new-state-space-sys-closed-loop} 
parametrized with respect to the matrix $G_{\mathcal{K}}$.

(ii) The parametrization \eqref{cl.dynamic} of the closed-loop system 
is the output-feedback counterpart of the parametrization \eqref{important3} 
obtained for the case of full state measurements. 
We can then proceed analogously to the proof of Theorem \ref{prop:data-driven-design} 
replacing $G_K, X_{0,T}, X_{1,T}$ with $G_{\mathcal{K}}, \hat X_{0,T}, \hat  X_{1,T}$ 
and obtain the claimed result {\it mutatis mutandis}. \qedp
\smallskip

Note that given a solution $\mathcal K$ as in 
\eqref{data-control-explicit-of} the resulting entries ordered 
as in \eqref{cal.K} lead to the following state-space realization
of order $n$ for the controller
\be\label{observer.form.w.inputs-controller}\ba{rl}
\xi(k+1) = &
\begin{bmatrix}
-c_n  & 1 & 0 & \cdots  & 0\\
-c_{n-1}  & 0 & 1 & \cdots &  0\\
\vdots & \vdots & \vdots & \ddots &  \vdots\\
0  & 0 & 0 & \cdots & 1\\
-c_1 & 0 & 0 & \cdots  & 0
\end{bmatrix}
\xi(k)
+
\begin{bmatrix}
d_n\\
d_{n-1}\\
\vdots \\
d_{2}\\
d_1
\end{bmatrix}
y(k)
\smallskip \smallskip \\ 
u(k)= &
\begin{bmatrix}
1  & 0 & 0 & \cdots & 0 & 0\\
\end{bmatrix}
\xi(k).
\ea
\ee

As a final point, we notice that
Theorem \ref{output-feedback-controller}
relies on the knowledge of the order $n$ of the system. In many cases, as for instance
in the numerical example which follows, this information can result from \emph{first principles} considerations. 
Otherwise, one can determine the model order from data, \emph{e.g.}
using subspace identification methods \cite[Theorem 2]{de_moor}. In this regard, it is worth pointing out 
that determining the model order from data does {not} correspond to
the whole algorithmic procedure needed to get a {parametric model} of the system.
Note that this information is also sufficient to render condition \eqref{rank.cond.of}
verifiable from data, which circumvents the problem of assessing persistence of excitation conditions 
that depend on the state trajectory of the system. 

\smallskip
\emph{Illustrative example}. Consider a system \cite{gahinet} made up by two carts.
The two carts are mechanically coupled by a
spring with uncertain stiffness $\gamma \in [0.25,1.5]$. 
The aim is to control the position of one cart by applying a force to the other cart. 
The system state-space description is given by 
\begin{eqnarray}
\left[
\begin{array}{c|c}
A & B \\ \hline C & D
\end{array}
\right]
=
\left[
\begin{array}{c|c}
\left [ \begin{array}{rrrr}  \, 0 \, & \, 1 \, & \, 0 \, & \, 0 \, \\ 
\, -\gamma \, & \, 0 \, & \, \gamma \, & \, 0 \, \\ \, 0 \, & \, 0 \, & \, 0 \, & \, 1 \, \\
\, \gamma \, & \, 0 \, & \, -\gamma \, & \, 0 \, \\ \end{array} \right ]
 & 
  \left [ \begin{array}{c}
 \,0 \, \\ \, 1 \, \\ \, 0 \, \\ \, 0 \, \end{array} \right ]
  \\ \hline 
   \left [ \begin{array}{cccc}
 \, \, 0 \,\, & \,\, 0 \,\, & \,\, 1 \,\, & \,\, 0 \,\, \end{array} \right ]
  & 0
\end{array}
\right].
\end{eqnarray}

Assume that $\gamma =1$ (unknown). 
{The system is controllable and observable. }
All the open-loop 
eigenvalues are on the imaginary axis. 
The input-output 
discretized version using a sampling time of $1s$
is as in \eqref{eq:difference} with coefficients 
\begin{eqnarray}
 \left[ 
\begin{array}{cccc}
a_1 & a_2 & a_3 & a_4
\end{array}
\right] = 
\left[ 
\begin{array}{cccc}
1 &   -2.311 &    2.623 &   -2.311
\end{array}
\right] \nonumber \smallskip \smallskip \\
 \left[ 
\begin{array}{cccc}
b_1 & b_2 & b_3 & b_4
\end{array}
\right] = 
\left[ 
\begin{array}{cccc}
0.039  &  0.383  &  0.383  &  0.039
\end{array}
\right]. \nonumber 
\end{eqnarray}
We design a controller following the approach described in 
Theorem \ref{output-feedback-controller}.
We generate the data with random initial conditions and by applying
a random input sequence of length $T=9$.
To solve 
\eqref{lmi.output.feedback}
we used CVX, obtaining from \eqref{data-control-explicit-of}
\begin{eqnarray} 
\mathcal K = 
\ba{rrrr}
 \big[ \,  1.1837 &  -1.5214  &  1.3408 & -1.4770  \\
    0.0005 &  -0.5035 &  -0.9589 & -0.9620 \, \big],
\ea \nonumber
\end{eqnarray}
which stabilizes the closed-loop dynamics in agreement with 
Theorem \ref{output-feedback-controller}. In particular, a minimal 
state-space representation $(A_c,B_c,D_c,D_c)$ 
of this controller is given by (see \eqref{observer.form.w.inputs-controller})
\begin{eqnarray}
\left[
\begin{array}{c|c}
A_c & B_c \\ \hline C_c & D_c
\end{array}
\right]
=
\left[
\begin{array}{c|c}
\left [ \begin{array}{rrrr}  \, -0.9620 \, & \, 1 \, & \, 0 \, & \, 0 \, \\ 
\, -0.9589 \, & \, 0 \, & \, 1 \, & \, 0 \, \\ \, -0.5035 \, & \, 0 \, & \, 0 \, & \, 1 \, \\
\, 0.0005 \, & \, 0 \, & \, 0 \, & \, 0 \, \\ \end{array} \right ]
 & 
  \left [ \begin{array}{r}
 \, -1.4770 \, \\ \, 1.3408 \, \\ \, -1.5214 \, \\ \, 1.1837 \, \end{array} \right ]
  \\ \hline 
   \left [ \begin{array}{cccc}
 \, \, 1 \,\, & \,\, 0 \,\, & \,\, 0 \,\, & \,\, 0 \,\, \end{array} \right ]
  & 0
\end{array}
\right]. \nonumber
\end{eqnarray}
\qedp \smallskip

\subsection{A remark on the case of MIMO systems} \label{sec:MIMO}

An analysis similar to the one presented before can be repeated 
starting from the left-difference operator of  a MIMO system,
\be\label{i/o.mimo}
\ba{r}
y(k) + A_{n} y(k-1) + \ldots + A_2 y(k-n+1) +A_1 y(k-n) \smallskip \\
= B_{n} u(k-1) 
+ \ldots + B_2 u(k-n+1) +B_1 u(k-n) \smallskip \smallskip
\ea\ee
where $y\in \mathbb{R}^p$, $u\in \mathbb{R}^m$, with $A_i$ and $B_i$
matrices of suitable dimensions. 
We define the state vector $\chi \in \mathbb{R}^{(m+p)n}$ as before 
which yields the  state representation \eqref{eq:new-state-space-sys-mimo}.
\begin{figure*}[!t]
\normalsize
{ 
{
\setcounter{MaxMatrixCols}{20}
\be\label{eq:new-state-space-sys-mimo}\ba{rl}
\chi(k+1) 
= &
\underbrace{
\begin{bmatrix}
0  & I_p & 0  & \cdots   & 0  
&
\mathbb{0}_{p\times m} & \mathbb{0}_{p\times m}  & \mathbb{0}_{p\times m} 
& \cdots 
& \mathbb{0}_{p\times m}
\\
0  & 0  & I_p & \cdots  & 0  
&
\mathbb{0}_{p\times m} & \mathbb{0}_{p\times m}  & \mathbb{0}_{p\times m} & \cdots &  \mathbb{0}_{p\times m}
\\
\vdots & \vdots &  \vdots  & \ddots & \vdots 
&
\vdots & \vdots &  \vdots  & \ddots &  \vdots 
\\
0  & 0  & 0  & \cdots &  I_p 
&
\mathbb{0}_{p\times m} & \mathbb{0}_{p\times m} & \mathbb{0}_{p\times m}  &  \cdots & \mathbb{0}_{p\times m}
\\
-A_1 &-A_2 & -A_3 & \cdots & -A_{n} &
B_1 &B_2 & B_3 & \cdots & B_{n} 
\\
\hline
\mathbb{0}_{m\times p} & \mathbb{0}_{m\times p}  & \mathbb{0}_{m\times p} & \cdots & \mathbb{0}_{m\times p} 
&
\mathbb{0}_{m\times m} & I_{m}  & \mathbb{0}_{m\times m} 
& \ldots 
& \mathbb{0}_{m\times m}
\\
\mathbb{0}_{m\times p} & \mathbb{0}_{m\times p}  & \mathbb{0}_{m\times p} & \cdots & \mathbb{0}_{m\times p} 
&
\mathbb{0}_{m\times m} &  \mathbb{0}_{m\times m}   & I_{m}
& \cdots 
& \mathbb{0}_{m\times m}
\\
\vdots & \vdots &  \vdots  & \ddots &  \vdots 
&
\vdots & \vdots &  \vdots  & \ddots &  \vdots 
\\
\mathbb{0}_{m\times p} & \mathbb{0}_{m\times p}  & \mathbb{0}_{m\times p} & \ldots & \mathbb{0}_{m\times p} 
&
\mathbb{0}_{m\times m} &  \mathbb{0}_{m\times m}   & \mathbb{0}_{m\times m}
& \ldots 
& I_{m}
\\
\mathbb{0}_{m\times p} & \mathbb{0}_{m\times p}  & \mathbb{0}_{m\times p} & \ldots & \mathbb{0}_{m\times p} 
&
\mathbb{0}_{m\times m} &  \mathbb{0}_{m\times m}   & \mathbb{0}_{m\times m}
& \ldots 
& \mathbb{0}_{m\times m}
\end{bmatrix}
}_{\mathcal{A}}
\chi(k)+
\underbrace{
\begin{bmatrix}
\mathbb{0}_{p\times m} 
\\
\mathbb{0}_{p\times m} 
\\
\vdots 
\\
\mathbb{0}_{p\times m} 
\\
\mathbb{0}_{p\times m} 
\\
\hline
\mathbb{0}_{m\times m} 
\\
\mathbb{0}_{m\times m} 
\\
\vdots 
\\
\mathbb{0}_{m\times m} 
\\
I_m\end{bmatrix}
}
_{\mathcal{B}}
u(k)
\\[2cm]
y(k) = & 
\underbrace{
\begin{bmatrix}
-A_1 &-A_2 & -A_3 & \cdots &  -A_{n} &
B_1 &B_2 & B_3 & \cdots & B_{n} 
\end{bmatrix}
}_{\mathcal{C}}
\chi(k)
\ea
\ee
}
}
\hrulefill
\vspace*{4pt}
\end{figure*}
In case of MIMO systems, we assume that 
we collect data with an input $u_{d, [0, T-1]}$, $T\ge ((m+p)n+1)(m+1)$,  
persistently exciting of order $(m+p)n+1$. 
Then, by Lemma \ref{lem:willems} we obtain the fulfilment of the following condition 
\be\label{full-row-rank-MIMO}
{\rm rank}
\begin{bmatrix}
U_{0,1,T} \\
\hline
\hat X_{0, T}
\end{bmatrix}
=
(m+p)n+m.
\ee
Under this condition, 
the same analysis of Section \ref{subsec:dd-of} can be repeated to obtain the following:
\smallskip

\begin{corollary}
Let condition \eqref{full-row-rank-MIMO} hold.
Then 
any matrix $\mathcal{Q}$ satisfying \eqref{lmi.output.feedback}
is such that the controller \be\label{eq:difference-control-mimo}
\ba{r}
y^c(k)+C_{n} y^c(k-1)+\ldots + C_{1} y^c(k-n) \smallskip \\
= 
D_{n} u^c(k-1)+\ldots + D_{1} u^c(k-n)
\ea\ee
with matrix coefficients given by 
\begin{eqnarray}
\label{cal.K.mimo}
\begin{bmatrix}
D_{1} & \ldots & D_{n} 
-C_{1} & \ldots & -C_{n} 
\end{bmatrix}
=
U_{0,1,T} \mathcal{Q} (\hat X_{0,T} \mathcal{Q} )^{-1}
\end{eqnarray}
stabilizes system \eqref{i/o.mimo}. 
Conversely, any controller as in \eqref{eq:difference-control-mimo}
that stabilizes \eqref{i/o.mimo} can be expressed in terms of the coefficients 
$\begin{bmatrix}
D_{1} & \ldots & D_{n} 
-C_{1} & \ldots & -C_{n} 
\end{bmatrix}
$
given by \eqref{cal.K.mimo}, with $\mathcal{Q}$ a solution to
\eqref{lmi.output.feedback}. 
\end{corollary}

\section{Discussion and conclusions} \label{sec:conclusion}
 
Persistently exciting data enable the construction of data-dependent 
matrices that can replace systems models. 
Adopting this paradigm proposed by \cite{willems2005note} 
we have shown the existence of  a parametrization of feedback control systems that allows us to reduce the stabilization problem to an equivalent data-dependent  linear matrix inequality. 
Since LMIs are ubiquitous in systems and control we expect that our approach could lead to data-driven  solutions to many other control problems. As an example we have considered an LQR problem. 
For several control problems, LMIs have proven their
effectiveness in providing robustness to various sources of uncertainty.
We have capitalized on this fact extending the analysis 
to the case of noise-corrupted data and showing  
how the approach can be used to stabilize unstable equilibria of nonlinear systems,
which are both situations where identification can be challenging.
A remarkable feature of all these results is that: (i) no parametric 
model of system is identified; (ii) stability guarantees come with a finite (computable) number of data points.

 
Studying how our approach can 
be used to systematically address control problems via 
data-dependent LMIs could be very rewarding, and lead 
to a methodical inclusion of data to analyze and design control systems. 
A great leap forward will come from systematically
extending the methods of this paper to systems where identification is challenging, such as switched \cite{Dai2018} and nonlinear systems. The results of this paper show that our approach 
is concretely promising for nonlinear systems, but we have only touched the surface of this research area.
Estimating the domain of attraction or considering other approaches such as \emph{lifting} techniques
are two simple examples of compelling research directions
for nonlinear systems.
Recent results have reignited  the interest of the community 
on system identification for nonlinear systems, interestingly pointing 
out the importance of the concept of persistently exciting signals 
\cite{padoan2015towards,padoan2017geometric}. We are confident  
that our approach will also play a fundamental role in developing 
a systematic methodology for the data-driven design of control laws 
for nonlinear systems. 

\appendix

\subsection{Proof of Lemma \ref{willems.fundamental}}

(i) By the Rouch\'{e}-Capelli theorem, the rank condition \eqref{rank.condition}  implies the existence of a vector $g\in \mathbb{R}^{T-t}$ such that 
\[
\bbm
u_{[0, t-1]}
\\
\hline
x_0
\ebm
=
\begin{bmatrix}
U_{0, t,T-t+1} \\
\hline
X_{0, T-t+1}
\end{bmatrix}
g.
\]
By replacing this expression in \eqref{fund.eqn}, we get
\[\ba{rl}
\bbm
u_{[0, t-1]}\\
y_{[0, t-1]}
\ebm
=&
\begin{bmatrix}
I_{tm} & \mathbb{0}_{tm\times n}\\
\hline
\mathcal{T}_t & \mathcal{O}_t
\end{bmatrix}
\begin{bmatrix}
U_{0, t,T-t+1} \\
\hline
X_{0, T-t+1}
\end{bmatrix}
g =
\begin{bmatrix}
U_{0,t,T-t+1}\\
\hline
Y_{0,t,T-t+1}
\end{bmatrix}
g
\ea\]
where the last identity holds because of \eqref{eq.Ht}. This concludes the proof of (i). 

(ii) In view of \eqref{eq.Ht}, 
\[
\begin{bmatrix}
U_{0,t,T-t+1}\\
\hline
Y_{0,t,T-t+1}
\end{bmatrix}g
=
\begin{bmatrix}
I_{tm} & \mathbb{0}_{tm\times t}\\
\hline
\mathcal{T}_t & \mathcal{O}_t
\end{bmatrix}
\begin{bmatrix}
U_{0,t,T-t+1} \\
\hline
X_{0, T-t+1}
\end{bmatrix}
g.
\]
Now define
\[
\bbm
u_{[0, t-1]}
\\
\hline
x_0
\ebm :=
\begin{bmatrix}
U_{0,t,T-t+1} \\
\hline
X_{0, T-t+1}
\end{bmatrix}
g
\]
Thus $U_{0,t,T-t+1}g$ represents a $t$-long input sequence $u_{[0, t-1]}$
of system \eqref{lin.sys}, while $Y_{0,t,T-t+1}g = \mathcal{O}_t x_0 + \mathcal{T}_t u_{[0, t-1]}$ 
is the corresponding output obtained from initial conditions $x_0$. \qedp

\subsection{Proof of Theorem \ref{prop:sys.identif}}

For compactness, let
\[
S := \begin{bmatrix}
U_{0,1,T} \\
\hline
X_{0, T}
\end{bmatrix}, \quad 
v :=
\begin{bmatrix}
u \\
x
\end{bmatrix}
\]
By the Rouch\'{e}-Capelli theorem, for any given $v$, the system 
of equations
\be\label{important}
v
=
S g
\ee
admits infinite solutions $g$, given by 
\be\label{g}
g= S^\dag
v
+ \Pi^\perp_S w, \quad  w\in \mathbb{R}^T,
\ee
where
$
\Pi^\perp_S
:=
\left(
I - S^\dag S
\right)
$
is the orthogonal projector onto the kernel of $S$.   
Hence,
\be\label{new.form}
x(k+1)
= 
\begin{bmatrix}
B & A
\end{bmatrix}
\begin{bmatrix}
u(k)\\
x(k)
\end{bmatrix} =
\begin{bmatrix}
B & A
\end{bmatrix}
S
g(k).
\ee
for some $g(k)$. 
As a final step, also note that
$ \begin{bmatrix} B & A
\end{bmatrix} S = X_{1, T}$. Overall, we thus have
\be\label{new.form.4}
x(k+1)
= 
X_{1, T}
\left(
S
^\dag
\begin{bmatrix}
u(k)\\
x(k)
\end{bmatrix}
+{\mathlarger{\Pi}}^\perp_{\tiny{
S
}
}
w(k)
\right)
\ee
with
$X_{1, T}
\Pi^\perp_S
=
\begin{bmatrix}
B & A
\end{bmatrix}
S \Pi^\perp_S
= 0$
where the last identity holds by the properties of the projector. \qedp

\subsection{Proof of Corollary \ref{prop:noise_stability_2}}

The idea for the proof is to show that Assumption \ref{ass:noise3} 
implies Assumption \ref{ass:noise2} with 
\begin{eqnarray}
\gamma = \frac{6 \gamma_1+3\gamma_2}{1 - 2 \gamma_1}
\label{4}
\end{eqnarray}
meaning that the proof
of Theorem \ref{prop:noise_stability} applies to Corollary \ref{prop:noise_stability_2}.

Suppose that \eqref{ass:noise3_a} holds.
By pre- and post-multiplying both terms of \eqref{ass:noise3_a}
by $[B \,\,\, A]$ and $[B \,\,\, A]^\top$
we get
{\setlength\arraycolsep{2pt} 
\begin{eqnarray}
&& A W_{0, T} W_{0, T}^\top A^\top \preceq \nonumber \\ 
&& \quad \gamma_1 (A Z_{0, T} + B U_{0,1,T}) (A Z_{0, T} + B U_{0,1,T})^\top =: \nonumber \\ 
&& \quad \gamma_1 V_{0, T} V_{0, T}^\top
\label{8}
\end{eqnarray}}%
where we set $V_{0, T} := A Z_{0, T} + B U_{0,1,T}$ for compactness.
Let us now write $\gamma_1$ as
{\setlength\arraycolsep{2pt} 
\begin{eqnarray}
\gamma_1 = \frac{\delta_1}{6 + 2\delta_1}
\quad \Longleftrightarrow \quad \delta_1 = \frac{6 \gamma_1}{1 - 2 \gamma_1}
\label{7}
\end{eqnarray}}%
Note that the above relation is well defined since $\gamma_1 \in (0,0.5)$ by hypothesis.
Also notice that for every $\gamma_1 \in (0,0.5)$ there uniquely corresponds $\delta_1>0$.

Hence, \eqref{8} can be rewritten as
{\setlength\arraycolsep{2pt} 
\begin{eqnarray}
\frac{3}{2}  A W_{0, T} W_{0, T}^\top A^\top \preceq 
\frac{\delta_1}{4} V_{0, T} V_{0, T}^\top
- \frac{\delta_1}{2} A W_{0, T} W_{0, T}^\top A^\top
\label{9}
\end{eqnarray}}%
Recall now that for arbitrary matrices $X,Y,F$ with $F \succ 0$, and  a scalar $\varepsilon > 0$, 
it holds that 
\begin{eqnarray}
X F Y^\top + Y F X^\top \preceq \varepsilon X F X^\top + \varepsilon^{-1} Y F Y^\top
\label{10}
\end{eqnarray}
By applying this property to the right hand side of \eqref{9} with $\varepsilon=0.5$, 
$X=V_{0, T}$, $F=I$ and $Y=A W_{0, T}$, we get
{\setlength\arraycolsep{2pt} 
\begin{eqnarray}
\frac{\delta_1}{4} V_{0, T} V_{0, T}^\top -
\frac{\delta_1}{2} A W_{0, T} W_{0, T}^\top A^\top &=& \nonumber \\
\frac{\delta_1}{2} \left[  V_{0, T} V_{0, T}^\top +
A W_{0, T} W_{0, T}^\top A^\top \right] && \nonumber \\
 \qquad -
 \frac{\delta_1}{2} 
 \left[ \frac{1}{2} V_{0, T} V_{0, T}^\top +
2 A W_{0, T} W_{0, T}^\top A^\top \right]
&\preceq& \nonumber \\
 \frac{\delta_1}{2} \left[ V_{0, T} V_{0, T}^\top +
A W_{0, T} W_{0, T}^\top A^\top \right] && \nonumber \\
 \qquad -
\frac{\delta_1}{2} \left[  A W_{0, T} V_{0, T}^\top +
V_{0, T} W_{0,T}^\top  A^\top \right] &=& 
\nonumber \\
 \frac{\delta_1}{2} \left[ ( V_{0, T} - A W_{0, T}) (V_{0, T} - A W_{0, T})^\top 
 \right]
&=& 
\nonumber \\
\frac{\delta_1}{2}  X_{1, T} X_{1, T}^\top
\end{eqnarray}}%
Thus \eqref{8} implies
{\setlength\arraycolsep{2pt} 
\begin{eqnarray}
\frac{3}{2}  A W_{0, T} W_{0, T}^\top A^\top \preceq 
\frac{\delta_1}{2} X_{1, T} X_{1, T}^\top
\label{12} 
\end{eqnarray}}%

Consider now \eqref{ass:noise3_b}, and 
let us write $\gamma_2$ as
{\setlength\arraycolsep{2pt} 
\begin{eqnarray}
\gamma_2 = \frac{\delta_2}{3 + \delta_1}
\quad \Longleftrightarrow \quad \delta_2 =  \gamma_2 (3 + \delta_1)
\end{eqnarray}}%
where $\delta_1$ has been defined in \eqref{7} and $\delta_2$ 
is a constant. 
Condition \eqref{ass:noise3_b} thus reads
{\setlength\arraycolsep{2pt} 
\begin{eqnarray}
3 W_{1, T} W_{1, T}^\top \preceq \delta_2 Z_{1, T} Z_{1, T}^\top - \delta_1 W_{1, T} W_{1, T}^\top
\label{14}
\end{eqnarray}}%
Combining \eqref{12} and \eqref{14} and using \eqref{10}, 
we finally verify that Assumption \ref{ass:noise2}  is 
satisfied with $\gamma$ as in \eqref{4}.
To see this, consider first the terms 
on the left hand side of \eqref{12} and \eqref{14}. By applying again (10) wit $\varepsilon =0.5$,
$X = A W_{0, T}$, $F=I$ and $Y=-W_{1, T}$ we obtain
{\setlength\arraycolsep{2pt} 
\begin{eqnarray}
R_{0, T}  R_{0, T}^\top =  \left(  A W_{0, T} - W_{1, T} \right)   \left(   A W_{0, T} - W_{1, T} \right)^\top
&\preceq& \nonumber \\ 
\frac{3}{2}  A W_{0, T} W_{0, T}^\top A^\top + 3 W_{1, T} W_{1, T}^\top
\end{eqnarray}}%
Consider next the terms 
on the right hand side of \eqref{12} and \eqref{14}. By applying again \eqref{10} with $\varepsilon =0.5$,
$X = X_{1, T}$, $F=I$ and $Y=-W_{1, T}$, we obtain
{\setlength\arraycolsep{2pt} 
\begin{eqnarray}
\frac{\delta_1}{2} X_{1, T} X_{1, T}^\top 
- \delta_1 W_{1, T} W_{1, T}^\top + \delta_2 Z_{1, T} Z_{1, T}^\top &=& \nonumber \\
 \delta_1 X_{1, T} X_{1, T}^\top +
\delta_1 W_{1, T} W_{1, T}^\top + \delta_2 Z_{1, T} Z_{1, T}^\top && \nonumber \\
- \delta_1 \left[ \frac{1}{2} X_{1, T} X_{1, T}^\top +
2 W_{1, T} W_{1, T}^\top \right] &\preceq& \nonumber \\
 \delta_1 X_{1, T} X_{1, T}^\top +
\delta_1 W_{1, T} W_{1, T}^\top + \delta_2 Z_{1, T} Z_{1, T}^\top && \nonumber \\
+\delta_1 X_{1, T} W_{1, T}^\top + \delta_1 W_{1, T} X_{1, T}^\top
&=& \nonumber \\
 \delta_1 ( X_{1, T} + W_{1, T}) ( X_{1, T} + W_{1, T})^\top +
\delta_2 Z_{1, T} Z_{1, T}^\top 
&=& \nonumber \\
 (\delta_1 + \delta_2 ) Z_{1, T} Z_{1, T}^\top 
 &=& \nonumber \\
\gamma Z_{1, T} Z_{1, T}^\top 
\end{eqnarray}}%
This gives the claim. \qedp

\bibliographystyle{IEEEtran}
\bibliography{biblio-data}

\end{document}